\documentclass[pra,aps,twocolumn,groupedaddress,superscriptaddress]{revtex4-2}
\usepackage[usenames,dvipsnames]{color}
\usepackage[colorlinks=true,linkcolor=Blue,citecolor=Blue,urlcolor=Blue]{hyperref}
\usepackage{graphicx}
\usepackage{bbm}
\usepackage{bm}
\usepackage{amsmath}
\usepackage{amssymb}
\usepackage{mathptmx}
\usepackage{hhline}
\usepackage[version=3]{mhchem}

\newcommand{\bs}{\bigskip}
\newcommand{\nn}{\nonumber \\}

\usepackage{array}
\usepackage[normalem]{ulem}

\usepackage{array}
\usepackage[normalem]{ulem}

\makeatletter
\newcommand{\thickhline}{%
    \noalign {\ifnum 0=`}\fi \hrule height 1pt
    \futurelet \reserved@a \@xhline
}
\newcolumntype{"}{@{\hskip\tabcolsep\vrule width 1pt\hskip\tabcolsep}}
\makeatother

\newcommand\redout{\bgroup\markoverwith{\textcolor{red}{\rule[.5ex]{2pt}{0.4pt}}}\ULon}
\newcommand\blueout{\bgroup\markoverwith{\textcolor{blue}{\rule[.5ex]{2pt}{0.4pt}}}\ULon}

\begin{document}
\title{Nested Shadows of Anyons:\\A Framework for Identifying Topological Phases}

\author{Yun-Tak \surname{Oh}}
\affiliation{Division of Semiconductor Physics, Korea University, Sejong 30019, Korea}

\author{Hyun-Yong \surname{Lee}}
\email{hyunyong@korea.ac.kr}
\affiliation{Division of Semiconductor Physics, Korea University, Sejong 30019, Korea}
\affiliation{Department of Applied Physics, Graduate School, Korea University, Sejong 30019, Korea}

\begin{abstract} 
The 1-form symmetries in two-dimensional topological systems are ``shadowed'' as global symmetries in their one-dimensional quantum transfer matrices. In this work, we introduce a distinct shadow effect arising from the pair-creation of anyons, which manifests as a local symmetry of the quantum transfer matrix. The interplay between these two shadow effects provides a powerful framework for characterizing topological phases without extensive numerical simulations. Specifically, we derive the phase diagram of the filtered toric code state and precisely identify phase boundaries using the nested shadows of anyons. Additionally, we reveal that a class of topological states host gapless edge modes protected by 1-form symmetry rather than global symmetry. Finally, we apply our approach to the three-dimensional toric code and X-cube states, uncovering a nontrivial path in phase space that connects them through a subdimensional critical point, which is highly challenging to detect numerically due to the complexity of simulating three-dimensional systems.
\end{abstract}

\date{\today}
\maketitle

\noindent \textbf{\large Introduction} \\
Identification of quantum phases of matter is a fundamental challenge in condensed matter physics. Conventional phases are distinguished by their order parameters, which characterize global symmetries of a system. In contrast, topological phases, characterized by non-local properties and robustness against local perturbations, defy the traditional symmetry-breaking paradigm\cite{Laughlin83, Wen90, Kane05, Balents10, Kane10, Wen17}. These phases are not described by local order parameters but rather by exotic entanglement patterns, making them fundamentally distinct from conventional phases of matter.

Topological phases encompass a wide range of phenomena, including symmetry-protected topological\,(SPT) phases, and intrinsic topological orders. SPT phases, protected by certain global symmetries, exhibit robust edge states that are robust to local perturbations unless the symmetry is explicitly broken\cite{chen12, senthil15}. The concept of SPT phases has been extended to systems with gapless bulk states\,\cite{Scaffidi17, Ruben21a, Ruben21b, Yu22, Yu24}. In particular, this has led to enriched quantum criticality through additional symmetry constraints\cite{Ruben21b}.  Intrinsic topological orders, such as those in fractional quantum Hall states and toric code, are characterized by long-range entanglement, anyonic excitations, and robust ground state degeneracies dependent on the system's topology\cite{Wen17}. 
The exploration of these exotic topological phases underscores their fundamental and practical importance in the broader field of condensed matter physics. In particular, it is becoming increasingly feasible to realize and simulate a wide variety of quantum phases with the advancement of quantum technologies\cite{google21, zhang22, shi23, Wang24, iqbal24}. In this context, a comprehensive understanding of topological phases, the mechanisms driving phase transitions, and the distinctions between these phases and other conventional phases is becoming essential. 

Nevertheless, the study of topological phases and their transitions is significantly hindered by the exponentially growing complexity of many-body quantum states, particularly in strongly interacting or higher-dimensional  systems. This computational challenge limits the ability to explore the full landscape of quantum phases. In light of this, tensor network\,(TN) representations, such as matrix product states\,(MPS) and projected-entangled pair states\,(PEPS), have emerged as a powerful framework for exploring topological phases. Beyond their computational utility, TNs also offer conceptual insights, enabling the classification of topological phases through the interplay of local symmetries and global structures\cite{Perez08,Pollmann10,Vidal10,Chen11,Schuch11,Andreas12,Pollmann12,Ilya17,Cirac21}. In particular, the gauge redundancies in the local tensors play a crucial role in describing the topological entanglement of the entire system, thereby enabling the characterization of intrinsic topological order. This gauge structure is generally represented by matrix product operators\,(MPOs) acting on the virtual indices of the TN, a property referred to as MPO-injectivity\cite{Bultinck17, Williamson17,Laurens21,Sahino21}. Many examples exhibit simpler forms of this gauge structure, the so-called $\mathbb{G}$-injective PEPS, where the MPO has a bond dimension of one. 
Furthermore, some of these examples might be referred to as 1-form symmetric PEPS\cite{He18, hy19, hy20, Tan24}, in which the action on the virtual indices is directly related to the physical index. This property allows one to identify the Wilson loop operator of the system explicitly. 

The zero-temperature partition function, or the norm of the ground state, $\langle \psi | \psi \rangle$, encodes information about the long-range physics, including spontaneous symmetry breaking (SSB) and topological characteristics of the system~\cite{haegeman15}. The dominant eigenvector and eigenvalue of transfer matrix $\mathbb{H}$ can be analyzed to compute the norm, expressed as:  
$ \langle \psi | \psi \rangle = \underset{L \rightarrow \infty}{\lim} {\rm Tr}(\mathbb{H}^L) = \underset{L \rightarrow \infty}{\lim} \lambda_0^L \langle 0 | 0 \rangle, $ where $L$ denotes the linear system size, $\lambda_0$ is the dominant eigenvalue, and $|0\rangle$ is the corresponding eigenvector representing the semi-infinite part of the norm. The TN representation of the wave function is highly effective for this purpose and has been extensively utilized and demonstrated to be valuable in previous studies~\cite{Schuch13, haegeman15, Norbert17, Schotte19, francuz20prb-1,francuz20prb-2, Iqbal21, chen21}. 

Despite significant progress, studies to date have primarily focused on analyzing $\mathbb{H}$ from the perspective of its global symmetries. However, in this work, we propose that an analysis based on the local symmetry of $\mathbb{H}$ is crucial for characterizing phase transitions and providing a refined classification of topological phases. 
First, we demonstrate that  quantum phases arising in variational wave functions and their transitions can be effectively identified by examining the interplay between global and local symmetries. 
Then, we demonstrate that local symmetry is not merely supplementary but essential for identifying a quantum phase, referred to as the {\bf 1-form symmetry-protected topological phase}, which cannot be fully characterized by global symmetries alone. Through this analysis, we establish that studying the local symmetry or the
projective representation of global symmetries of $\mathbb{H}$ is essential for identifying topological phases. 
The local symmetries are intrinsically related to the pair-creation processes of anyons, with their corresponding symmetry operators explicitly defined at the fixed points. Although the exact forms of these local symmetries become less transparent away from the fixed points, their existence remains evident, as they originate from the fundamental pair-creation operations of anyons. This suggests that our discussion applies broadly to tensor network representation of topologically ordered states.  
Finally, we extend this framework to challenging three-dimensional models, where numerical simulations encounter significant limitations. By analyzing only local tensors, we uncover a potential non-trivial pathway connecting the $3d$ toric code and X-cube states, which passes through a subdimensional critical point. \\

\noindent \textbf{\large Results} \\
\noindent \textbf{Global symmetries of transfer matrix}. 
We consider a translation-invariant quantum state of which wave function is represented by a uniform PEPS as follows:
\begin{align}
    \includegraphics[width=0.8\linewidth]{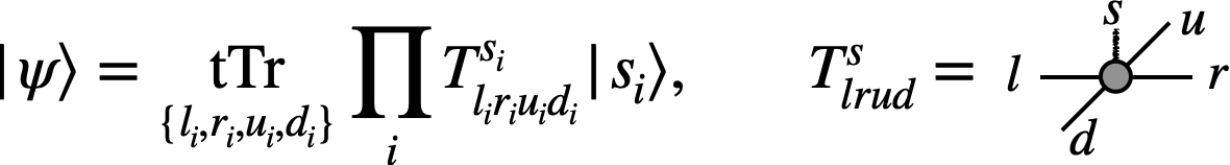}
\end{align}
where ${\rm tTr}_{\{\cdot\}}$ denotes the tensor trace over the virtual indices $\{\cdot\}$, and $s_i$ stands for the local quantum number at site $i$. We denote the dimension of the virtual indices as $D$.
\begin{figure}[t!]
    \centering
    \includegraphics[width=0.99\linewidth]{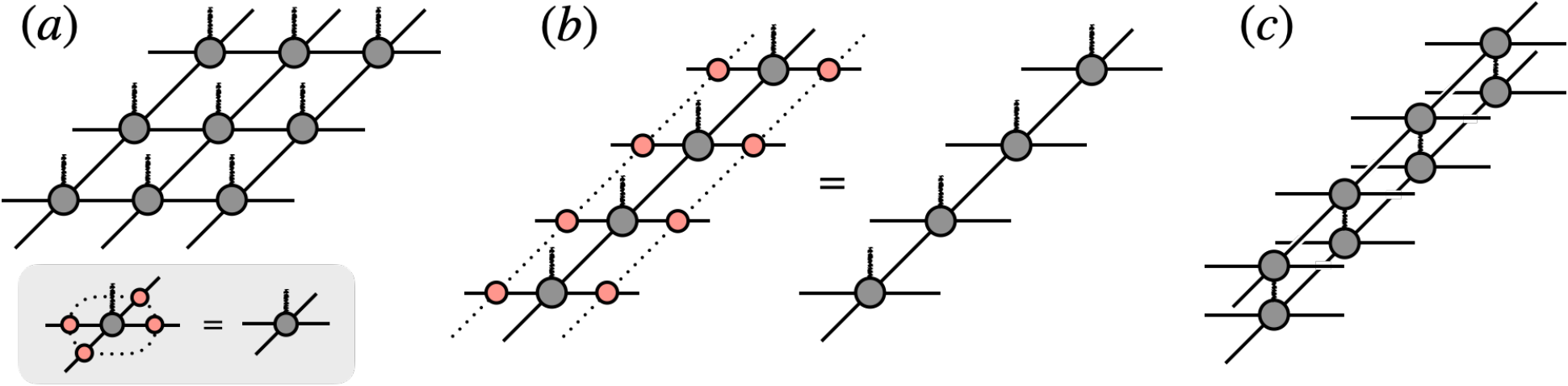}
    \caption{Schematic illustrations of (a) an injective PEPS on a square lattice, (b) the invariance of a column of local tensors under a gauge transformation, and (c) the transfer matrix for the PEPS. }
    \label{fig:schematic}
\end{figure}
The schematic illustration of PEPS is shown in Fig.\,\ref{fig:schematic}\,(a). We further assume that the PEPS is injective, guaranteeing the existence of gauge transformations that leave the local \( T \)-tensor invariant, as illustrated in Fig.\,\ref{fig:schematic}\,(a). Specifically, for a $\mathbb{G}$-injective PEPS, this invariance is expressed as $ g_{ll'} g_{uu'} g^{-1}_{r'r} g^{-1}_{d'd} T_{l'r'u'd'}^s = T_{lrud}^s $ where the gauge transformation $g$ forms the group $\mathbb{G}$. In this work, we focus on the $\mathbb{G}$-injective PEPS for simplicity; however, the protocol we develop can be generalized to the MPO-injective PEPS. Now, we define a TN patch consisting of a column of $T$-tensors:
\begin{align}
    \mathbb{C}_{\{l_i\}, \{r_i\}}^{\{s_i\}} \equiv \underset{\{u_i,d_i\}}{\rm tTr} \prod_{i\in {\rm column}} T_{l_i r_i u_i d_i}^{s_i}, \nonumber
\end{align}
or $\mathbb{C}$ in short. Using the gauge invariance of the $T$-tensor, one can show that the $\mathbb{C}$-TN is invariant under transformations applied to the open virtual indices: ${\bm g} \mathbb{C} {\bm g}^{-1} = \mathbb{C}$ with ${\bm g} \equiv g^{\otimes{\rm column}}$ as depicted graphically in Fig.\,\ref{fig:schematic}\,(b). Consequently, the transfer matrix\,(TM), defined as  
\begin{align}
    \mathbb{H}^{\{\overline{l}_i\},\{\overline{r}_i\}}_{\{l_i\},\{ r_i\}} \equiv {\rm tTr}_{\{s_i\}} \mathbb{C}_{\{l_i\}, \{r_i\}}^{\{s_i\}} \mathbb{C}_{\{\overline{l}_i\}, \{\overline{r}_i\}}^{*\{s_i\}},
    \nonumber
\end{align}
and graphically represented in Fig.\,\ref{fig:schematic}\,(c), remains invariant under the transformations \({\bm g} \otimes \overline{{\bm g}}'\) with \(g, g' \in \mathbb{G}\): 
\(
    ({\bm g} \otimes \overline{\bm g}')\, \mathbb{H}\, ({\bm g} \otimes \overline{\bm g}')^{-1} = \mathbb{H}.
\)
Here, the operator with overline, $\overline{o}$, acts on the virtual indices in the upper\,(bra) layer. This indicates that \(\mathbb{H}\) can be interpreted as a one-dimensional quantum Hamiltonian with a local Hilbert space of dimension $D\times D$, which preserves a global $\mathbb{G} \times \mathbb{G}$-symmetry. Even though it originates from the gauge redundancy of the local $T$-tensor, the global symmetry of the TM provides valuable information for understanding the quantum phases described by the PEPS. Furthermore, its SSB have been utilized to analyze the long-range properties of the state and the fate of anyonic excitations, e.g., their condensation and confinement/deconfinement\,\cite{Schuch13, haegeman15, Norbert17, Schotte19, Iqbal21}. To detect SSB and characterize phase boundaries, one may employ the one-dimensional tensor network solvers such as the variational uniform matrix product states to compute the dominant eigenstate of the TM $\mathbb{H}$. \\

\noindent \textbf{Local symmetries of transfer matrix.} 
Having additional knowledge of local symmetry offers significant advantages beyond what global symmetry alone can provide. First, the fate of certain global symmetries can be determined without large-scale simulations, as local symmetry may dictate them itself. 
More intriguingly, local symmetry enables the characterization of exotic phases beyond Landau's paradigm, such as SPT phase. 
Here, we derive the local symmetries of $\mathbb{H}$ and then present their invaluable utility with an exemplary PEPS ansatz extensively discussed in existing literature.

\begin{figure}[t!]
    \centering
    \includegraphics[width=0.99\linewidth]{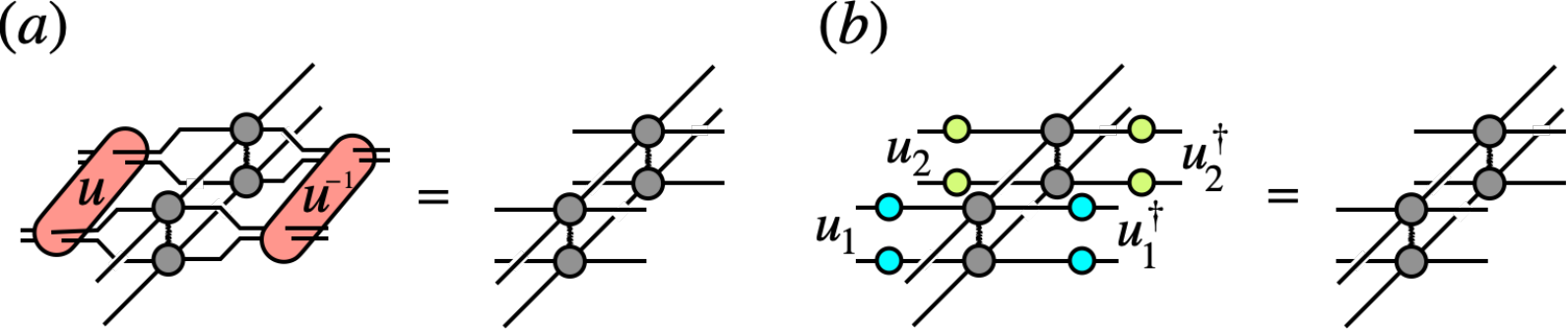}
    \caption{Graphical representations of (a) the invariance of the transfer matrix under a two-site general local similarity transformation and (b) a special case where the transformation is a direct product of operations on each virtual index. }
    \label{fig:local_symmetry}
\end{figure}

Local symmetry of the TM can be demonstrated by invertible transformations ${\bm u}_i$ acting only on a finite patch of $\mathbb{H}$ around site $i$, i.e., ${\bm u}_i \mathbb{H} {\bm u}_i^{-1} = \mathbb{H}$.  Figure\,\ref{fig:local_symmetry}\,(a) illustrates the local transformation acting on a two-site patch of $\mathbb{H}$. The local symmetries discussed in this study are represented by unitary operators acting on single-site or two-site patches in a tensor product form, \({\bm u}_i = \prod_{j\in i} (u_j \otimes \overline{u}_j^*)\), where \(u_j\) operates solely on a single virtual index \(l_j\), as schematically illustrated in Fig.\,\ref{fig:local_symmetry}\,(b). However, our argument naturally extends to more general cases where local symmetry is realized through entangled operators acting on multi-site patches.\\

\noindent \textbf{Toric code wave function.} 
We begin with the toric code\,(TC) state, whose wave function can be efficiently represented as a uniform PEPS on a square lattice. The corresponding $T$-tensor, $T_{lrud}^{s_1 s_2}$, includes two physical  and four virtual indices of $D=2$. See Supplementary Note for explicit definition of the $T$-tensor. The tensor exhibits a gauge redundancy, $Z_{ll'} Z_{uu'} Z_{r'r} Z_{d'd} T_{l'r'u'd'}^{s_1 s_2} = T_{lrud}^{s_1 s_2}$, where $Z$ represents the Pauli matrix. Therefore, the TM of the TC state can be regarded as an $S=1/2$ quantum spin model preserving the $\mathbb{Z}_2 \times \mathbb{Z}_2$ global symmetry, generated by ${\bm g}_{ZI} \equiv \prod_i (Z_i \otimes \overline{I}_i)$ and ${\bm g}_{ZZ} \equiv \prod_i (Z_i \otimes \overline{Z}_i)$.

We now note that the TM, $\mathbb{H}^{\rm TC}$, is also invariant under two local symmetries, \({\bm u}_i^z = Z_i \otimes \overline{Z}_i\) and \({\bm u}_i^x = X_i \otimes \overline{X}_i\) for all \(i\). See Supplementary Note for a detailed derivation.
Note that the two local symmetries commute with each other. Combined with Elitzur's theorem, this implies that the physical states of $\mathbb{H}^{\rm TC}$ must be simultaneous eigenstates of both ${\bm u}_i^z$ and ${\bm u}_i^x$. Given that ${\bm g}_{ZZ} = \prod_i {\bm u}_i^z$ and $\{ {\bm g}_{ZI}, {\bm u}_i^x \} = 0$, the global ${\bm g}_{ZZ}$ symmetry is preserved by ${\bm u}_i^z$, while ${\bm g}_{ZI}$ is spontaneously broken by ${\bm u}_i^x$ in all eigenstates.
Consequently, the TM eigenstates are direct product of local Bell states, indicating that the TC state is a trivial fixed-point state with zero correlation length.
\\

\noindent \textbf{Phase diagram of the filtered toric code.} 
We now examine a deformation of the TC state through filtering operations: 
$|{\rm fTC}(h_x, h_z)\rangle \equiv \prod_i \left( 1 + h_x X_i + h_z Z_i\right) |{\rm TC}\rangle.$
The phase diagram of the filtered TC\,(fTC) state has been extensively studied~\cite{castelnovo08, haegeman15prx, chen21}. Here, we discuss the utility of local symmetries and their interplay with global symmetries in mapping out potential phases and characterizing the nature of phase boundaries, without relying on simulations.

To begin, note that the filtering operation is an action of direct product of local operators, and therefore does not affect the injectivity of the TC state. This ensures that its TM, $\mathbb{H}_{(h_z, h_x)}^{\rm fTC}$, commutes with both ${\bm g}_{ZI}$ and ${\bm g}_{ZZ}$ at arbitrary $(h_z, h_x)$. In contrast, the $Z\,(X)$-filtering operation preserves only the local symmetry ${\bm u}_i^{z\,(x)}$:  $[{\bm u}_i^z,\, \mathbb{H}^{\rm fTC}_{(h_z, 0)}]=0$ and $[{\bm u}_i^x,\, \mathbb{H}^{\rm fTC}_{(0, h_x)}]=0$. This implies that the global \({\bm g}_{ZZ}\) symmetry remains preserved in \(\mathbb{H}_{(h_z, 0)}^{\rm fTC}\), whereas \({\bm g}_{ZI}\) is always broken in \(\mathbb{H}_{(0, h_x)}^{\rm fTC}\). Consequently, the \(Z\)-filtering operation induces at most a single phase transition, corresponding to the restoration of \({\bm g}_{ZI}\). On the other hand, the breaking of \({\bm g}_{ZZ}\) is the only possible transition in the \(X\)-filtering operation.

We can pinpoint the occurrence of phase transitions by analyzing the local symmetries at two points. First, at \((h_z, h_x) = (1,0)\), the TM becomes symmetric under local transformations \(Z_i \otimes \overline{I}_i\) and \(I_i \otimes \overline{Z}_i\). These local symmetries guarantee the restoration of \({\bm g}_{ZI}\) in the \(Z\)-filtering operation, \(\langle {\bm g}_{ZI} \rangle = 1\), indicating that the phase transition must occur at or below \(h_z = 1\), i.e., \(h_z^c \leq 1\). While the exact critical point is not located, the phase diagram is consistent with the one obtained from numerical simulations~\cite{Zhu19}. Similarly, two local symmetries, \(X_i \otimes \overline{I}_i\) and \(I_i \otimes \overline{X}_i\), emerge at \((h_z, h_x) = (0,1)\), ensuring the breaking of the \({\bm g}_{ZZ}\) symmetry, $\langle {\bm g}_{ZZ}\rangle = 0$. This also provides the upper bound for the critical point in the \(X\)-filtering operation, $h_x^c \leq 1$. Since both transitions involve the \(\mathbb{Z}_2\) symmetry breaking, the critical points are described by the \((1+1)d\) Ising conformal field theory (CFT) with central charge $c=1/2$.

\begin{figure}[t!]
    \centering
    \includegraphics[width=0.99\linewidth]{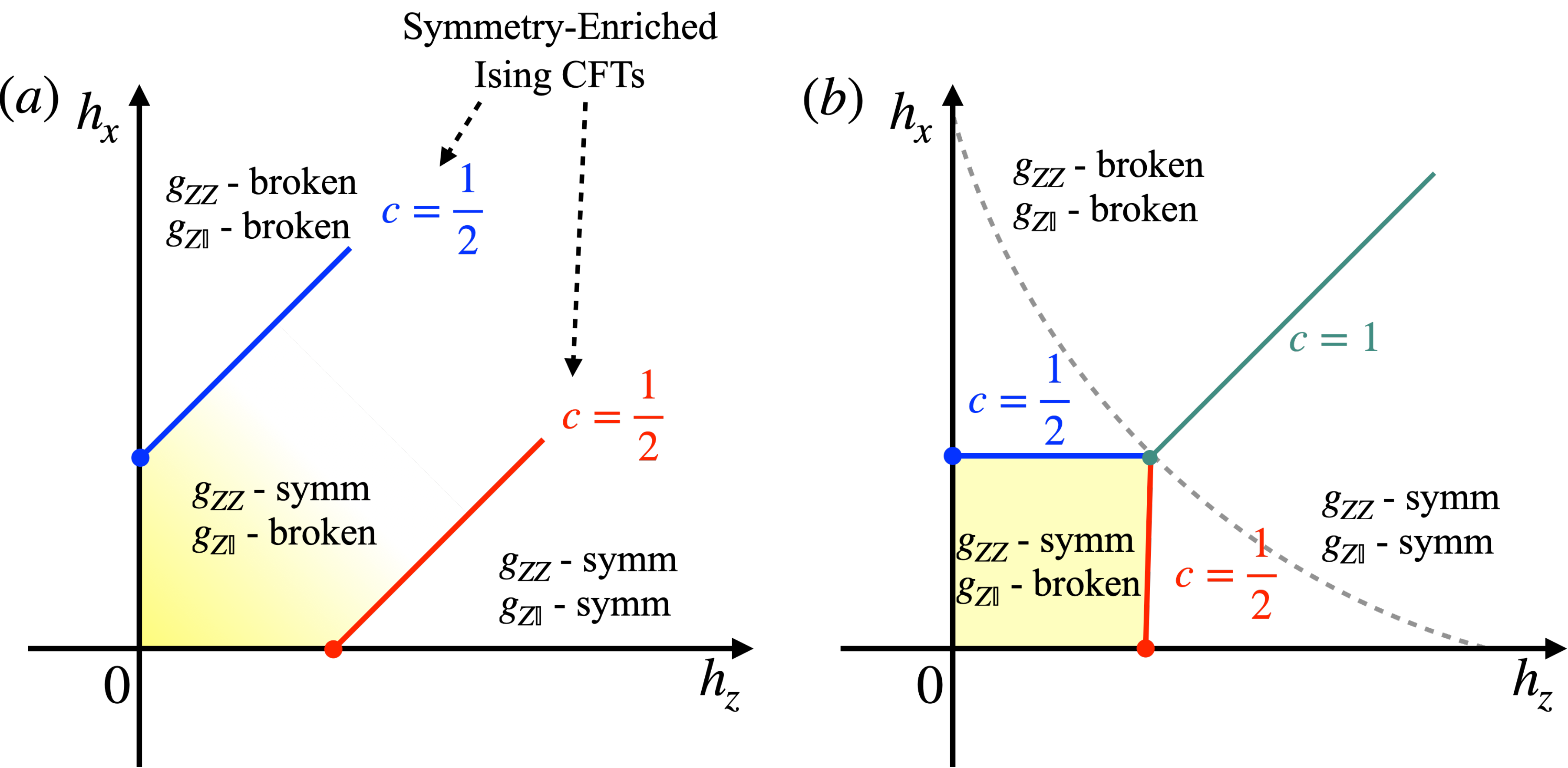}
    \caption{Phase diagram of the filtered TC. (a) The phase diagram is constructed exclusively based on the interplay between local and global symmetries. It highlights three phases with distinct breaking patterns of the global ${\bm g}_{ZZ} \times {\bm g}_{ZI}$ symmetries, along with two symmetry-enriched Ising CFT lines that separate them. (b) The phase diagram is enhanced by the inclusion of additional information regarding the global symmetry present along the dashed curve. At a point on this curve, the two distinct Ising CFTs merge into the $({\rm Ising})^2$ CFT with $c=1$.}
    \label{fig:dtc_phase_diagram}
\end{figure}
\begin{table}[b!]
\centering
    \begin{tabular}{|c||c|c|c|c|}
    \hline
    ansatz & $G_{\rm eff}$ & $G_{\rm gap}$ & 
    $O_i$ & model \cite{Ruben21a}
    \\ \thickhline 
    $X$-filtered TC\,($h_x^c$) & ${\bm g}_{ZZ}$ & $\mathbb{I}$ &  $Z_i\otimes \overline{Z}_i$ & $H_0 + H_x$ \\ \hline
    $Z$-filtered TC\,($h_z^c$) & ${\bm g}_{ZZ}\times {\bm g}_{ZI}$         & ${\bm g}_{ZZ}$  & $Z_i\otimes \overline{Z}_i$  & $H_1 + H_x$   \\ \hline
    LG state\,($\theta_c$)  & ${\bm g}_{ZZ}\times {\bm g}_{ZI}$         & ${\bm g}_{ZZ}$  & $v_i\otimes \overline{v}_i^\dagger$  & $H_{\rm Hal} + H_y$\\ \hline
    \end{tabular}
    \caption{ The unbroken symmetry $G_{\rm eff}$, gapped symmetry $G_{\rm gap}$, and the end-point of string operator $O_i$ of the transfer matrix, $\mathbb{H}$, for each ansatz at the critical point. The fourth column specifies the model classified in Ref.\,\cite{Ruben21a}, which belongs to the same class of symmetry-enriched Ising CFT. }
    \label{tab:ising_cft}
\end{table}

By analyzing the effective symmetry group \(G_{\rm eff}\) and the gapped symmetry group \(G_{\rm gap}\), we further identify that the two critical points belong to two distinct classes among the nine different \(\mathbb{Z}_2 \times \mathbb{Z}_2\)-enriched Ising CFTs\,\cite{Ruben21b}, as summarized in Table\,\ref{tab:ising_cft}. Therefore, the two critical points cannot be smoothly connected in the parameter space $(h_z, h_x)$, preserving both \({\bm g}_{ZZ}\) and \({\bm g}_{ZI}\) symmetries. Instead, if they intersect, the transition must involve an additional critical point characterized by a \(c=1\) CFT\,\cite{Ruben21b}.

\begin{figure}[t!]
    \centering
    \includegraphics[width=1\linewidth]{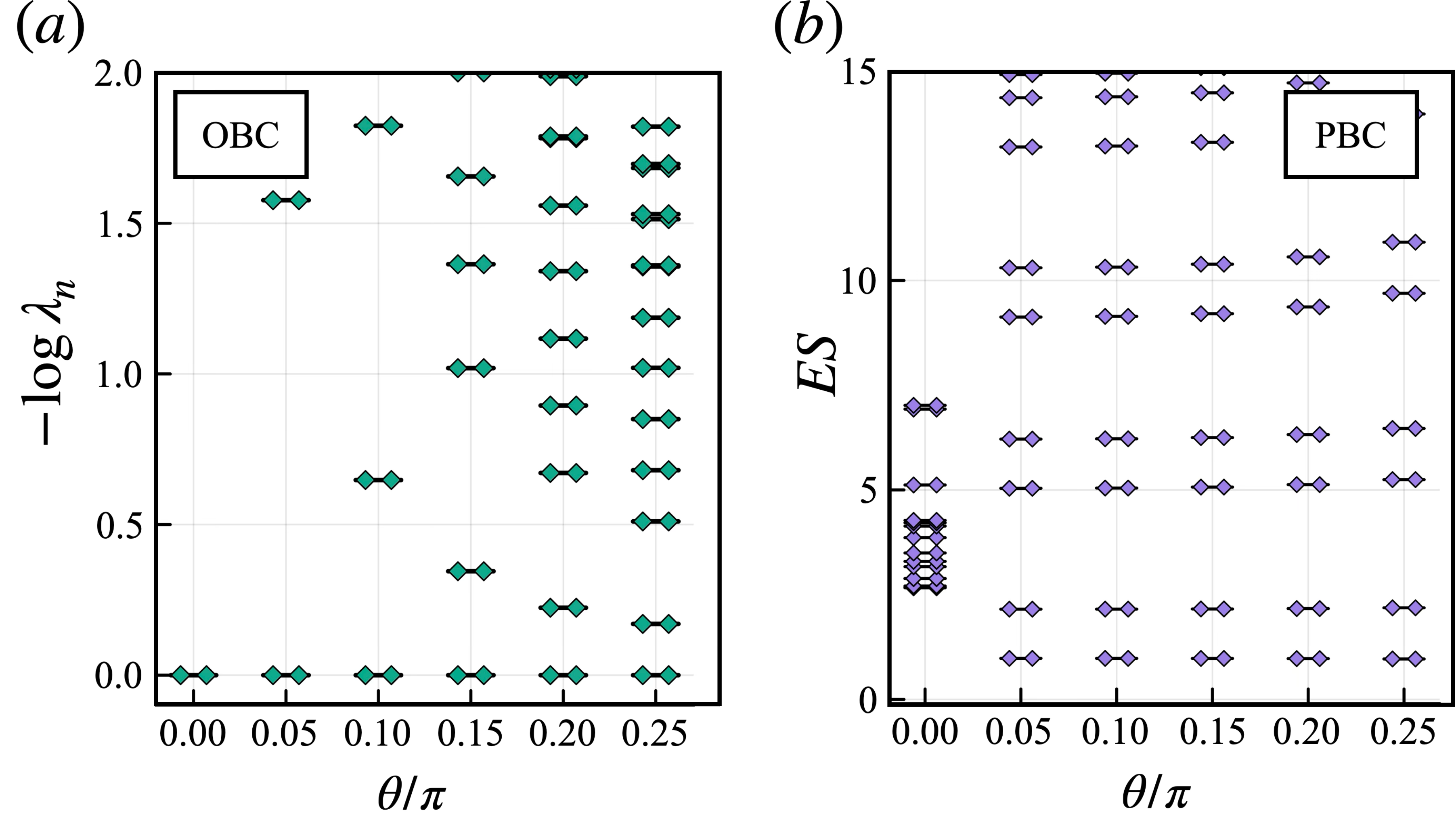}
    \caption{
    (a) Eigenvalue spectrum of \(\mathbb{H}^{\rm LG}_{\theta}\) under OBC.  (b) Entanglement spectrum of the dominant eigenstate of \(\mathbb{H}^{\rm LG}_{\theta}\) under PBC. The system size is fixed to \(L=10\).  In both (a) and (b), a clear double degeneracy appears for entire range of \(\theta \).
    }
    \label{fig:lg_dtc_spectrum}
\end{figure}

The phase diagram shown in Fig.\,\ref{fig:dtc_phase_diagram}(a) is constructed \textbf{exclusively based on the interplay between the local and global symmetries} of  \(\mathbb{H}_{(h_z, h_x)}^{\rm fTC}\). In the Supplementary Note, we demonstrated the existence of a curve along which an additional accidental global symmetry emerges, causing both \({\bm g}_{ZZ}\) and \({\bm g}_{ZI}\) to be either simultaneously preserved or simultaneously broken along this curve. This curve is indicated by a dashed line in Fig.\,\ref{fig:dtc_phase_diagram}(b). Consequently, the two symmetry-enriched Ising CFTs meet at a point on this curve as shown in Fig.\,\ref{fig:dtc_phase_diagram}(b). This critical point is commonly referred to as $({\rm Ising})^2$, which corresponds to a $c=1$ orbifold CFT~\cite{Ruben21b}. The critical point is known to extend along the \(h_z = h_x\) line~\cite{Zhu19}, representing the direct transition between the fully symmetric and broken phases. This is the only aspect that cannot be determined through local symmetry analysis. Nonetheless, it is remarkable how long-range physics, such as SSB and the nature of phase transitions, can be readily determined from the properties of the local \(T\)-tensor or a finite patch of the TN, without relying on large-scale simulations.

It is important to emphasize that at the TC point, \((h_z,h_x) = (0,0)\), applying \({\bm u}_i^z \) or \({\bm u}_i^x \) creates a pair of either \(m\)- or \(e\)-anyons near the virtual index $i$ in both ket and bra layers.
In other words, these two local symmetries originate from the fact that \(m\)- and \(e\)-anyons are well-defined in the quantum state. On the other hand, at the points \((h_z, h_x) = (1,0)\) and \((h_z, h_x) = (0,1)\), the \(m\)-anyon and \(e\)-anyon, respectively, are condensed. In these cases, the pair-creation of anyons on the quantum state effectively acts as an trivial operation, which is reflected in the emergence of local symmetries in the TM: \(Z_i\otimes \overline{I}_i\) or \(I_i\otimes \overline{Z}_i\) for \(m\)-anyon condensation, and \(X_i\otimes \overline{I}_i\) or \(I_i\otimes \overline{X}_i\) for \(e\)-anyon condensation. Since these local symmetries originate from pair-creation processes of anyons—which are inherently `local'—such local symmetries are expected be present in the TM of any topologically ordered state. Ultimately, the phase of the fTC state at any point is determined by the interplay between the local and global symmetries of \(\mathbb{H}^{\rm fTC}_{(h_z, h_x)}\). As a result, we believe that the argument regarding the interplay between local and global symmetries remains universally valid for any topologically ordered state.  
\\

\noindent \textbf{Kitaev honeycomb loop gas state.} 
We now demonstrate that the local symmetry of the TM serves as a powerful tool for identifying quantum phases that are indistinguishable by global symmetry considerations alone. Let us consider the so-called loop gas\,(LG) ansatz, \( |{\rm LG}(\theta)\rangle \), which was introduced for the Kitaev honeycomb model (KHM) in Ref.\,\cite{hy19}. By grouping two sites within the unit cell into a single site, the LG state can be represented as a $D=2$ PEPS on the square lattice. See the Supplementary Note for the explicit definition of the \(T\)-tensor and the variational parameter \(\theta\).
One can verify that the $T$-tensor is also invariant under the gauge transformation: $Z_{ll'} Z_{uu'} Z_{r'r} Z_{d'd} T_{l'r'u'd'}^{s_1 s_2} = T_{lrud}^{s_1 s_2}$. Therefore, the TM of the LG state, $\mathbb{H}_{\theta}^{\rm LG}$, also enjoys the global ${\bm g}_{ZI} \times {\bm g}_{ZZ}$ symmetry. 

In \(0\leq \theta \leq \pi/4\), two points allow for exact solutions. At \(\theta=0\), the ansatz corresponds to the exact ground state of the KHM in the strong anisotropic limit\,\cite{hy21}, representing a \(\mathbb{Z}_2\) topologically ordered state\,\cite{kitaev06}. In contrast, at \(\theta_c = \pi/4\), it becomes a critical state characterized by \((1+1)d\) Ising CFT\,\cite{hy19}. Therefore, the LG state in this parameter range  closely resembles that of the \(Z\)-filtered TC state in \(0\leq h_z \leq h_z^c\). Moreover, numerical simulations (without careful consideration) identify a dominant ground state of \(\mathbb{H}_\theta^{\rm LG}\) that breaks the \({\bm g}_{ZI}\) symmetry. In this context, \(\mathbb{H}_{(h_z,0)}^{\rm fTC}\) and \(\mathbb{H}_{\theta}^{\rm LG}\) appear indistinguishable in terms of global symmetries, suggesting a smooth deformation between them. However, remarkably, we find that these two TMs are topologically distinct, and the local symmetry is a crucial key to reveal this distinction.

To explore this, we first note that \(\mathbb{H}_\theta^{\rm LG}\) remains invariant under a two-site local transformation 
\begin{align}
  {\bm v}_i \equiv ( v_{i-1}^* \otimes \overline{v}_{i-1})(v_i \otimes \overline{v}^*_i)  
\end{align}
%
with \(v = \frac{1}{2}(1+i)(X\!-\!Y)\), i.e., \({\bm v}_i^\dagger \mathbb{H}^{\rm LG}_{\theta} {\bm v}_i = \mathbb{H}^{\rm LG}_{\theta}\) for any \(\theta\). This local symmetry enforces \(\langle X_i \otimes \overline{I}_i\rangle = 0\), ensuring the preservation of \({\bm g}_{ZZ}\) across all \(\theta\). Furthermore, it induces long-range string order in the eigenstates of \(\mathbb{H}_\theta^{\rm LG}\), given by  
\(
\lim_{|n-m|\rightarrow \infty}\left\langle S_m^\dagger S_n \right\rangle = 1,
\)
where the string operator  
\begin{align}
    S_n \equiv \prod_{i\leq n} {\bm v}_{i} = \left(\prod_{i<n} Z_{i} \otimes \overline{Z}_{i} \right)(v_n \otimes \overline{v}^*_n).
\end{align}
Here, \(vv^* = iZ\) is used. Notably, this acts as a symmetry flux of ${\bm g}_{ZZ}$ with a non-trivial end-point \( v \otimes \overline{v}^* \), akin to the string-operator with the end-point Majorana operator in the Kitaev chain model. It is well established that long-range string order not only induces  gapless edge modes\,\cite{kennedy92, Pollmann12, Pollmann12b, Ruben21b}. To probe this characteristic, we diagonalize \(\mathbb{H}^{\rm LG}_{\theta}\) for finite systems as a function of \(\theta\). The eigenvalue spectrum exhibits a twofold degeneracy for all \(\theta\) under open boundary conditions (OBC) as illustrated in Fig.\,\ref{fig:lg_dtc_spectrum}(a). 
In fact, the breaking of \({\bm g}_{ZI}\) under OBC observed in simulations arises from the arbitrary superposition of two degenerate eigenstates. In a \({\bm g}_{ZI}\)-symmetric basis, the degenerate eigenstates are related through end-point operators, i.e., \( |1\rangle = v_1 \otimes v^*_1 |0\rangle = v^*_L \otimes v_L |0\rangle \). In contrast, under periodic boundary conditions (PBC), the dominant eigenvector is unique, yet its bipartite entanglement spectrum retains a twofold degeneracy. Overall, \(\mathbb{H}_{\theta\leq \theta_c}^{\rm LG}\) exhibits a topological phase closely resembling that of the Kitaev chain. Consequently, we conclude that \(\mathbb{H}_{\theta\leq \theta_c}^{\rm LG}\) belongs to an SPT phase protected by ${\bm g}_{ZI}$, rendering it topologically distinct from any phase in the fTC state. We also stress that, instead of global symmetry, it is the system’s {\it 1-form symmetry} that protects the topological edge states in the LG state.

The string operator provides a sharper characterization of the critical point as well. Since \({\bm g}_{ZZ}\) is preserved for all \(\theta\), the effective symmetry group and the gapped symmetry group at \(\theta_c\) are given by \(G_{\rm eff} = {\bm g}_{ZZ} \times {\bm g}_{ZI}\) and \(G_{\rm gap} = {\bm g}_{ZZ}\), respectively. Therefore, \(G_{\rm eff}\) and \(G_{\rm gap}\) do not differentiate this critical point from that of the \(Z\)-filtered TC\,[See Table\,\ref{tab:ising_cft}].  
However, the charge of the end-point operator under \({\bm g}_{ZI}\) provides a crucial distinction. In the case of the \(Z\)-filtered TC, the string operator \(\prod_{i\leq n}{\bm u}_i^z = \prod_{i\leq n} Z_i\otimes \overline{Z}_i\) has a trivial end-point under \({\bm g}_{ZI}\). In contrast, in the LG state, the end-point operator is nontrivial, satisfying \({\bm g}_{ZI} (v_i \otimes \overline{v}_i^*){\bm g}_{ZI} = - (v_i \otimes \overline{v}_i^*)\). This discrete distinction indicates that the two critical points belong to distinct symmetry-enriched Ising criticalities\,\cite{Ruben21a}.

\begin{figure}[t!]
    \centering
    \includegraphics[width=0.7\linewidth]{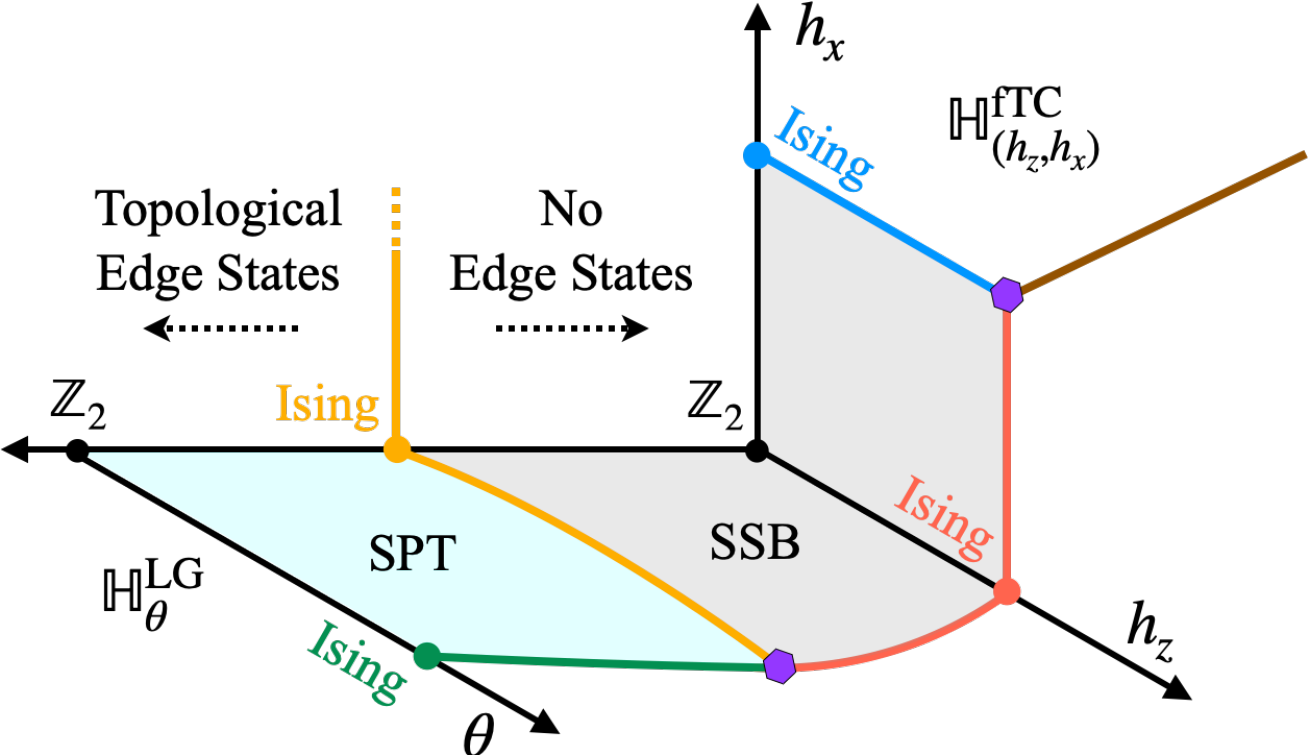}
    \caption{Schematic phase diagram of the dominant eigenvalue of the transfer matrices $\mathbb{H}_\theta^{\rm LG}$ and $\mathbb{H}_{(h_z,h_x)}^{\rm fTC}$. The colored solid lines represent different symmetry-enriched \((1+1)d\) Ising CFTs\,[See Table.\,\ref{tab:ising_cft}], while multi-critical points are indicated by purple hexagons. The dominant eigenvector of \(\mathbb{H}_{(h_z,h_x)}^{\rm fTC}\) in the toric code phase is a trivial direct product state breaking the ${\bm g}_{ZI}$ symmetry, whereas that of \(\mathbb{H}_\theta^{\rm LG}\) hosts gapless edge modes protected by the \({\bm g}_{ZI}\) symmetry. }
    \label{fig:lg_dtc_phase_diagram}
\end{figure}

It is also noteworthy that a SSB of \({\bm g}_{ZI}\) must be involved in deforming the LG state at \(\theta=0\) into the TC phase in the fTC state, despite both describing the same topological order. In fact, in the strong anisotropic limit, the KHM maps to Wen’s plaquette model, which, when translation symmetry is considered, belongs to a distinct symmetry-enriched topologically ordered (SET) phase from the TC model. The distinction between these two SET phases lies in the presence of Majorana edge modes~\cite{lu16prb}, aligning with our result. 
Summarizing the phases of the fTC and LG states, the overall phase diagram is depicted in Fig.\,\ref{fig:lg_dtc_phase_diagram}. Our study emphasizes that the projective representation of global symmetry in the TM serves as a crucial tool for characterizing topological phases that host gapless edge modes. Furthermore, we believe that our approach provides a framework for distinguishing different SET phases, achieving a level of classification that a global symmetry analysis alone cannot. The mathematical formulation of this framework is left for future investigation.
\\

\noindent \textbf{$3d$ Toric code and X-cube states.} 
Now, we discuss the local symmetry in the TM of three-dimensional quantum states and its utilization, focusing on the $3d$ TC and X-cube states. These states can be efficiently represented by $3d$ PEPS on a cubic lattice, with $T$-tensors having three physical indices. The TMs are then represented by two-dimensional projected entangled-pair operators\,(PEPO), and can be interpreted as $(2+1)d$ quantum models. See Supplementary Note for the definition of the $T$-tensors and detailed derivations of the global and local symmetries discussed below.

By extending the discussion in the $2d$ TC state, it is straightforward to show that the TMs of both states, $\mathbb{H}^{\rm 3TC}$ and $\mathbb{H}^{\rm XC}$, are invariant under the global symmetries ${\bm g}_{ZZ}$ and ${\bm g}_{ZI}$, as well as the local symmetries ${\bm u}_i^z$ and ${\bm u}_i^x$. Consequently, the dominant eigenvectors are the identical product state, $|0\rangle$, that break the ${\bm g}_{ZI}$ symmetry, mirroring the $2d$ TC state: $u_i^z |0\rangle = |0\rangle$ and $ u_i^x |0\rangle = |0\rangle$.
In what follows, we explore two possible paths in Hilbert space that preserve both ${\bm g}_{ZZ}$ and ${\bm g}_{ZI}$, connecting $\mathbb{H}^{\rm 3TC}$ and $\mathbb{H}^{\rm XC}$ by leveraging their local properties.

First, we consider a rather straightforward deformation: the $Z$-filtering operation applied to both states, i.e., $\prod_i (1+h_z X_i) |{\rm 3TC}\rangle$ and $\prod_i (1+h_z X_i) |{\rm XC}\rangle$.
Similar to the $2d$ fTC case, the $Z$-filtering operation preserves ${\bm g}_{ZZ}$ due to ${\bm u}_i^z$ but induces a phase transition involving the SSB of ${\bm g}_{ZI}$, characterized by the $(2+1)d$ Ising universality class in both cases. The upper bound of the critical point is given by $h_z^c \leq 1$, which can be determined from the emergent local symmetries at $h_z = 1$. This implies the existence of a path in the parameter space connecting $3d$ TC and XC states, which passes through two $(2+1)d$ Ising critical points.

\begin{figure}[t!]
    \centering
    \includegraphics[width=0.9\linewidth]{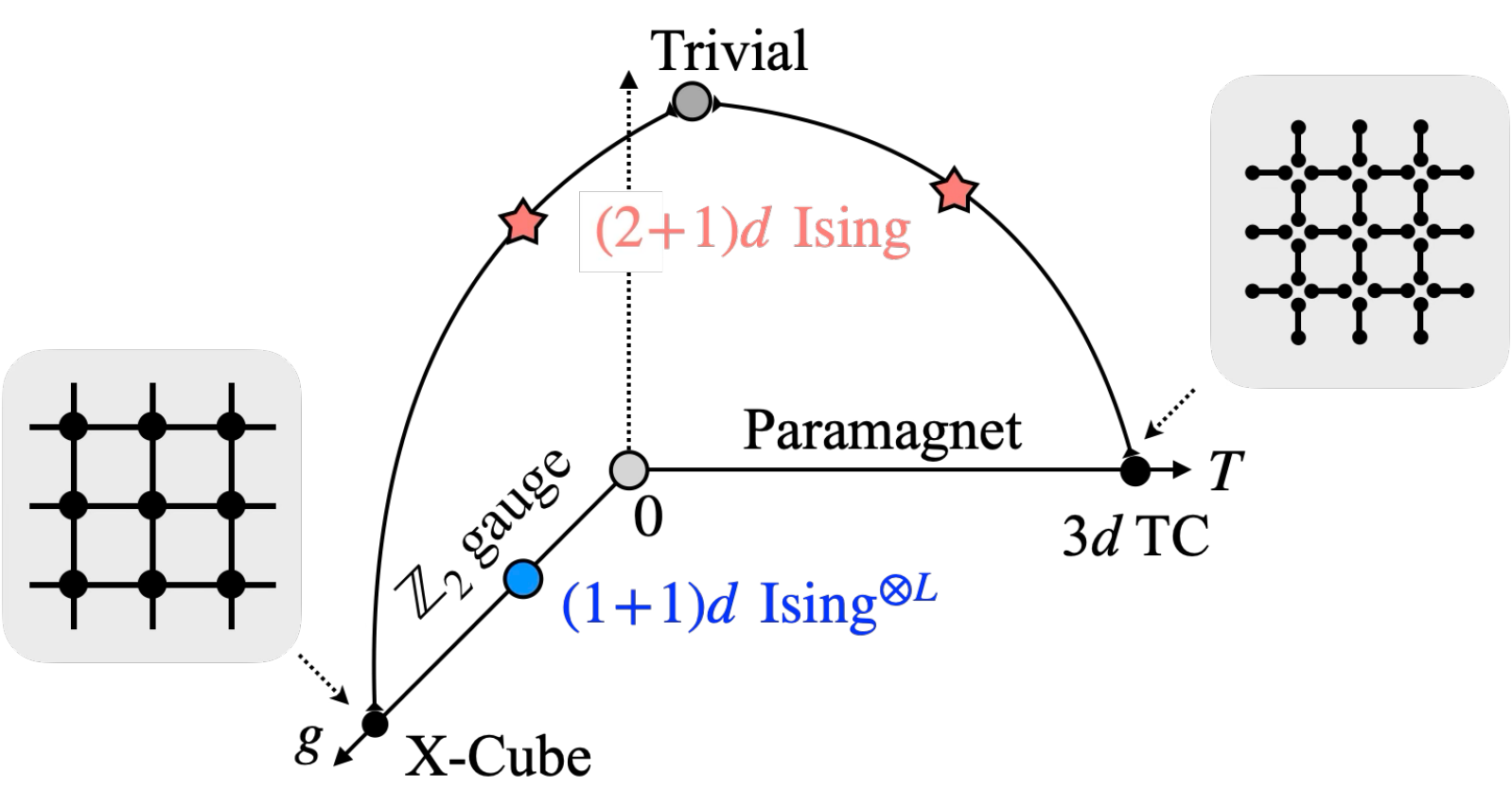}
    \caption{Schematic phase diagram illustrating two possible paths connecting the \(3d\) TC state and the X-cube state, while preserving the \(\mathbb{Z}_2\) gauge symmetry of the wavefunction. The dominant eigenvalue of the transfer matrix for the \(3d\) TC state corresponds to the classical partition function of a paramagnet in the infinite-temperature limit. Each eigenvalue can be represented by a two-dimensional TN, as depicted in the gray-shaded square.
    In contrast, for the X-cube state, it corresponds to the \(\mathbb{Z}_2\) gauge theory on a square lattice in the infinite coupling constant (\(g\)) limit. See text for details.}
    \label{fig:3dtc_xcube}   
\end{figure}

Then, we identify a non-trivial path connecting these two states. By examining local properties of the TMs, we find that \(\mathbb{H}^{\rm 3TC}\) and \(\mathbb{H}^{\rm XC}\) can be expressed as idempotent operators, given by:
\begin{align}
    \mathbb{H}^{\rm 3TC} &= \prod_{\langle ij\rangle} \left[1+ (X_i \otimes \overline{X}_i)(X_j \otimes \overline{X}_j) \right], \nonumber\\
    \mathbb{H}^{\rm XC} &= \prod_{\square} \Big[1+ \prod_{i\in \square} (X_i \otimes \overline{X}_i)\Big],
\end{align}
where $\langle ij\rangle$ denotes neighboring sites, and $i\in\square$ represents sites on an elementary plaquette in the square lattice. The above equalities can be verified within a finite patch of the TMs, similar to how local symmetries are analyzed. See Supplementary Note for detailed derivations. Consequently, while the eigenvectors are identical, the eigenvalues are distinct:
$ \mathbb{H}^{\rm 3TC} |0\rangle = Z_{\rm Para}^{T=\infty} |0\rangle \quad \text{and} \quad \mathbb{H}^{\rm XC} |0\rangle = Z_{\mathbb{Z}_2}^{g=\infty} |0\rangle$. Here, $Z_{\rm Para}^T$ denotes the  classical partition functions of a trivial paramagnet, $H_{\rm Para} = - \sum_i Z_i$ at temperature $T$, while $Z_{\mathbb{Z}_2}^{g}$  stands for the zero-temperature partition function of the $\mathbb{Z}_2$ gauge theory on square lattice, $H_{\mathbb{Z}_2} = - g\sum_\square \prod_{i\in \square} X_i - g^{-1} \sum_i Z_i $ at coupling constant $g$. As a result, the norm of two states are given by:  
\begin{align}
  \langle 3{\rm TC} | 3{\rm TC} \rangle = \left[ Z_{\rm Para}^{T=\infty} \right]^L, \quad
  \langle {\rm XC} | {\rm XC} \rangle = \left[Z_{\mathbb{Z}_2}^{g=\infty}\right]^L,
\end{align}
where $L$ represents the number of layers or slices in the $3d$ system. Since the $\mathbb{Z}_2$ gauge theory exhibits the $(1+1)d$ Ising critical point describing the confinement/deconfinement transition, one can envision a deformation of the XC state such that the eigenvalue, $Z_{\mathbb{Z}_2}^g$, changes continuously from $g=\infty$ to $g=0$. Then, the deformation must involve a phase transition characterized by a stacking of $(1+1)d$ Ising CFTs. We do not find such a deformation with $D=2$ PEPS, but we cannot completely rule out its existence in an enlarged PEPS Hilbert space with larger bond dimensions. 
Similarly, one can imagine a deformation of the $3d$ TC state that continuously changes the eigenvalue $Z_{\rm Para}^T$ to its zero-temperature limit. Along this deformation, no phase transition occurs. This implies that, since $Z_{\mathbb{Z}_2}^{g=0} = Z_{\rm Para}^{T=0}$, the XC and $3d$ TC states can be transformed into each other along this path undergoing the subdimensional criticality\,\cite{Lake21, Brandon23}. We present a schematic phase diagram in Fig.\,\ref{fig:3dtc_xcube}, along with the TN representation of the eigenvalues. 
\\

\noindent \textbf{\large Conclusion} \\
We have demonstrated that analyzing the local symmetry of the transfer matrix, combined with its interplay with global symmetry, provides a powerful framework for distinguishing quantum phases and characterizing the nature of phase transitions without relying on numerical simulations or classical partition function analysis. Remarkably, even though long-range information is typically required to characterize features such as SSB, we have shown that these properties can be inferred from the characteristics of a local tensor or a finite patch of TN. 

Furthermore, we have shown that local symmetry plays an essential role in distinguishing topological phases, far beyond simply serving as a supplementary tool for completing phase diagrams. By leveraging local symmetry, we have identified the 1-form symmetry-protected topological phase, a quantum phase that cannot be determined by global symmetry or classical partition function analysis. This novel phase exhibits topological edge states protected by the 1-form symmetry rather than the global symmetry of the system. 

Finally, we have extended this framework to challenging three-dimensional models, where numerical simulations face severe limitations. By focusing on the properties of local tensors, we have proposed possible quantum phase transitions, including sub-dimensional critical point, between $3d$ toric code and X-cube states. These transitions are challenging to determine through numerical approaches; however, we have inferred them from local tensor analysis. This demonstrates the applicability and effectiveness of our method in exploring higher-dimensional systems and complex quantum phases that are otherwise inaccessible with conventional methods.

Since the local symmetries of the transfer matrix stem from the local pair-creation of anyons, our approach—leveraging the interplay between local and global symmetries—is expected to be broadly applicable to topologically ordered states across different SET phases. By providing a new perspective on these symmetries, our method extends existing approaches for exploring fundamental questions related to anyon condensation and confinement~\cite{haegeman15}.
\\

\noindent \textbf{\large Code availability} \\
Exact diagonalization codes in this paper are available from the authors upon reasonable request. \\

\noindent \textbf{\large Data availability}\\
All relevant data in this paper are available from the authors upon reasonable request. \\

\noindent \textbf{\large Acknowledgement} \\
Y.-T.O. is grateful to J. Kim for the many valuable discussions. 
Y.-T.O. acknowledges support from the National Research Foundation of Korea (NRF) under Grant No. NRF-2022R1I1A1A01065149. H.-Y.L. and Y.-T.O. were supported by the NRF grant funded by MSIT
under Grants No. RS-2023-00220471. \\

\noindent \textbf{\large Author contributions} \\
H.-Y.L. conceived the project and supervised the study. Y.-T.O. performed the tensor network and exact diagonalization computations. Y.-T.O. and H.-Y.L contributed to the writing of the manuscript. \\

\noindent \textbf{\large Competing interests} \\
The authors declare no competing interests. \\

\noindent \textbf{\large Additional information} \\
\textbf{Supplementary information} is available for this paper. \\

\bibliography{references}

\clearpage
\bs 

\pagebreak
\bs 

\onecolumngrid
\newpage
\begin{center}

\bs 
\bs 
\textbf{\large Supplementary Material for ``Nested Shadows of Anyons: A Framework for Identifying Topological Phases''}\\[.3cm]
\end{center}

\setcounter{section}{0}
\setcounter{equation}{0}
\setcounter{figure}{0}
\setcounter{table}{0}
\setcounter{page}{1}

\bs

\section{Toric Code Loop Gas Operator}
%
\subsection{Gauge Symmetry and Physical-Virtual Relation}

The projected entangled pair state (PEPS) representation of the \(\mathbb{Z}_2\) toric code (TC) can be efficiently represented as a tensor product with bond dimension \(D = 2\):
\begin{align}
|\psi_{\rm TC}\rangle = {\rm tTr} \prod_{\alpha} T_{u_{\alpha}d_{\alpha}l_{\alpha}r_{\alpha}}^{\tilde{s}_\alpha}|\tilde{s}_\alpha\rangle.
\end{align}
Here, \(\tilde{s} = \{s^x, s^y\}\) represents the local physical degrees of freedom. The local tensor \(T_{udlr}^{\tilde{s}}\) can be decomposed into two types of smaller tensors, as illustrated below:
\begin{align}
    \includegraphics[width=0.5\linewidth]{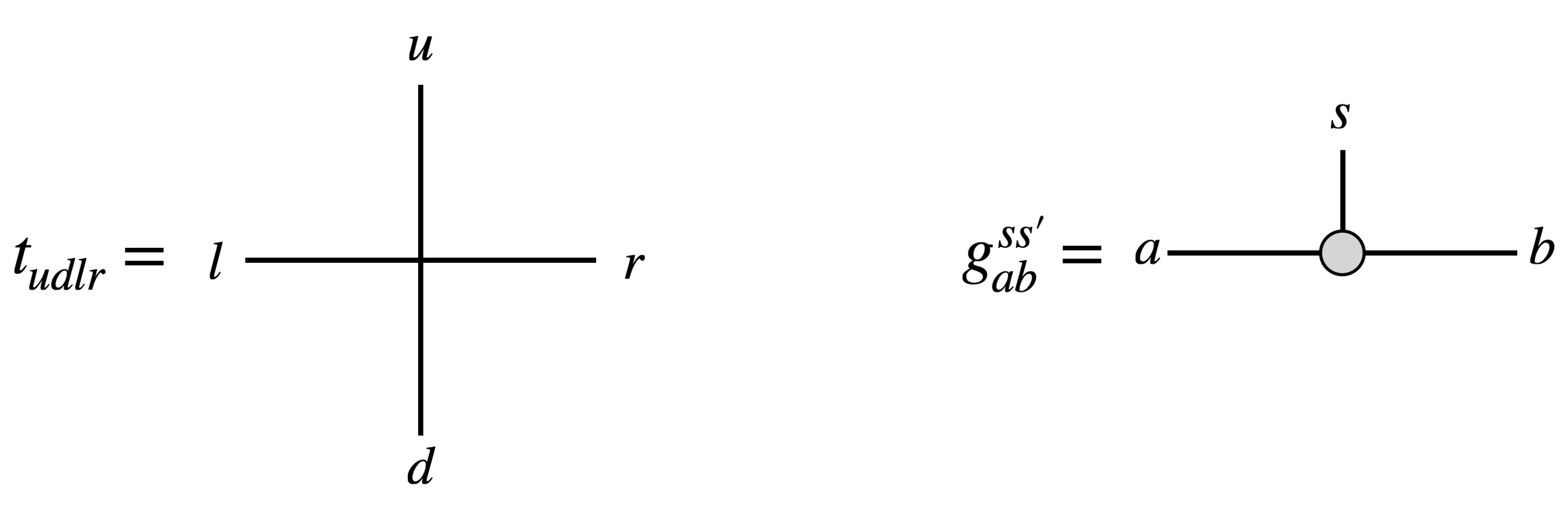}
    \label{eq:tc-small-tn}
\end{align}
These tensors are defined as:
\begin{align}
t_{udlr} &= 1 \quad \text{if and only if}~ (u-d+r-l)\,{\rm mod}\,2 = 0,
\nn
& = 0 \quad \text{otherwise,} 
\end{align}
and
\begin{align}
g_{ab}^{s} & = 1 \quad \text{if and only if}~ a = b = s,
\nn
& = 0 \quad \text{otherwise.}
\label{eq:g_tensor}
\end{align}
Notably, the index \(s=0\) corresponds to the local physical state \(|0\rangle = (1,0)\), while \(s=1\) corresponds to \(|1\rangle = (0,1)\).

The local tensor \(T_{udlr}^{\tilde{s}}\) can then be expressed as \(T_{udlr}^{\tilde{s}} = \sum_{u'r'} t_{u'dlr'} g_{r'r}^{s_x} g_{u'u}^{s_y}\). This relationship can also be represented graphically:
\begin{align}
    \includegraphics[width=0.38\linewidth]{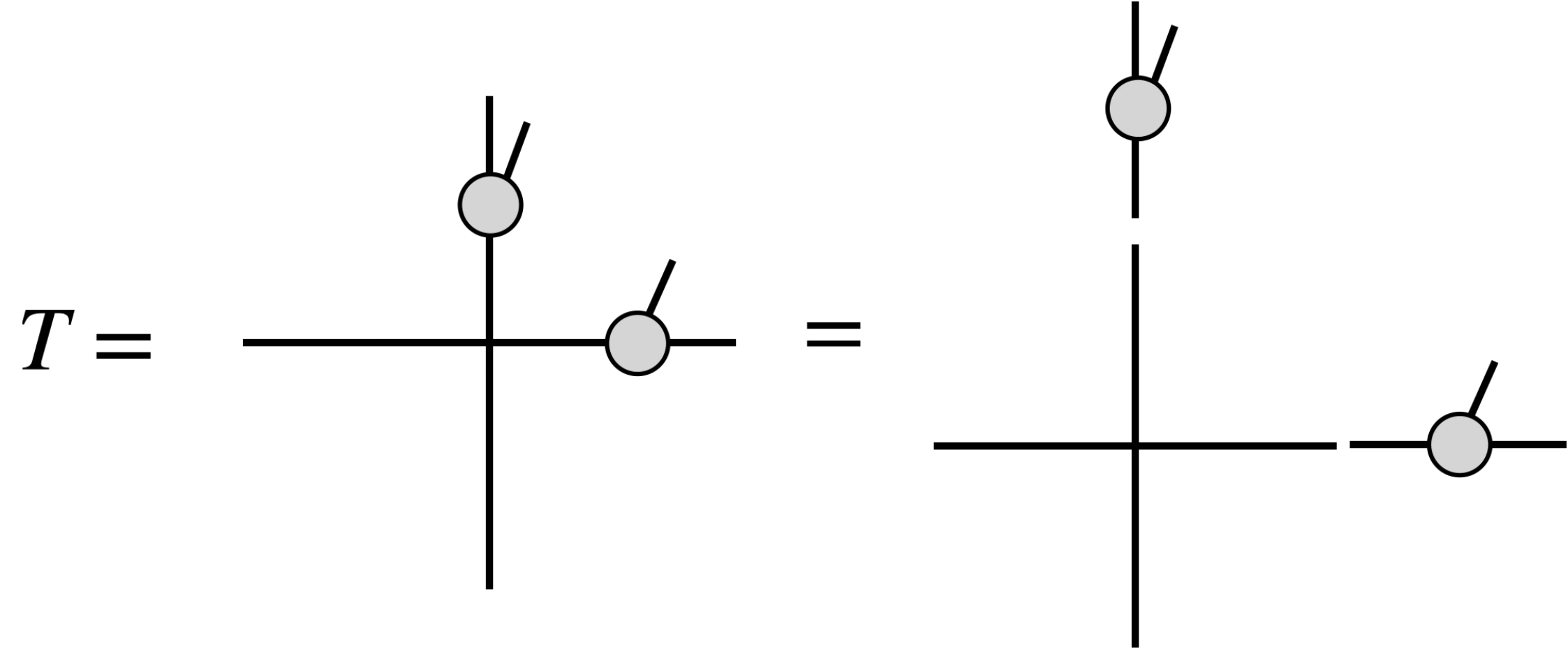}
    \label{eq:tc-T}
\end{align}
The smaller tensors \(g^{s}_{ab}\) and \(t_{udlr}\) exhibit the following local gauge symmetries:
\begin{align}
    \includegraphics[width=0.45\linewidth]{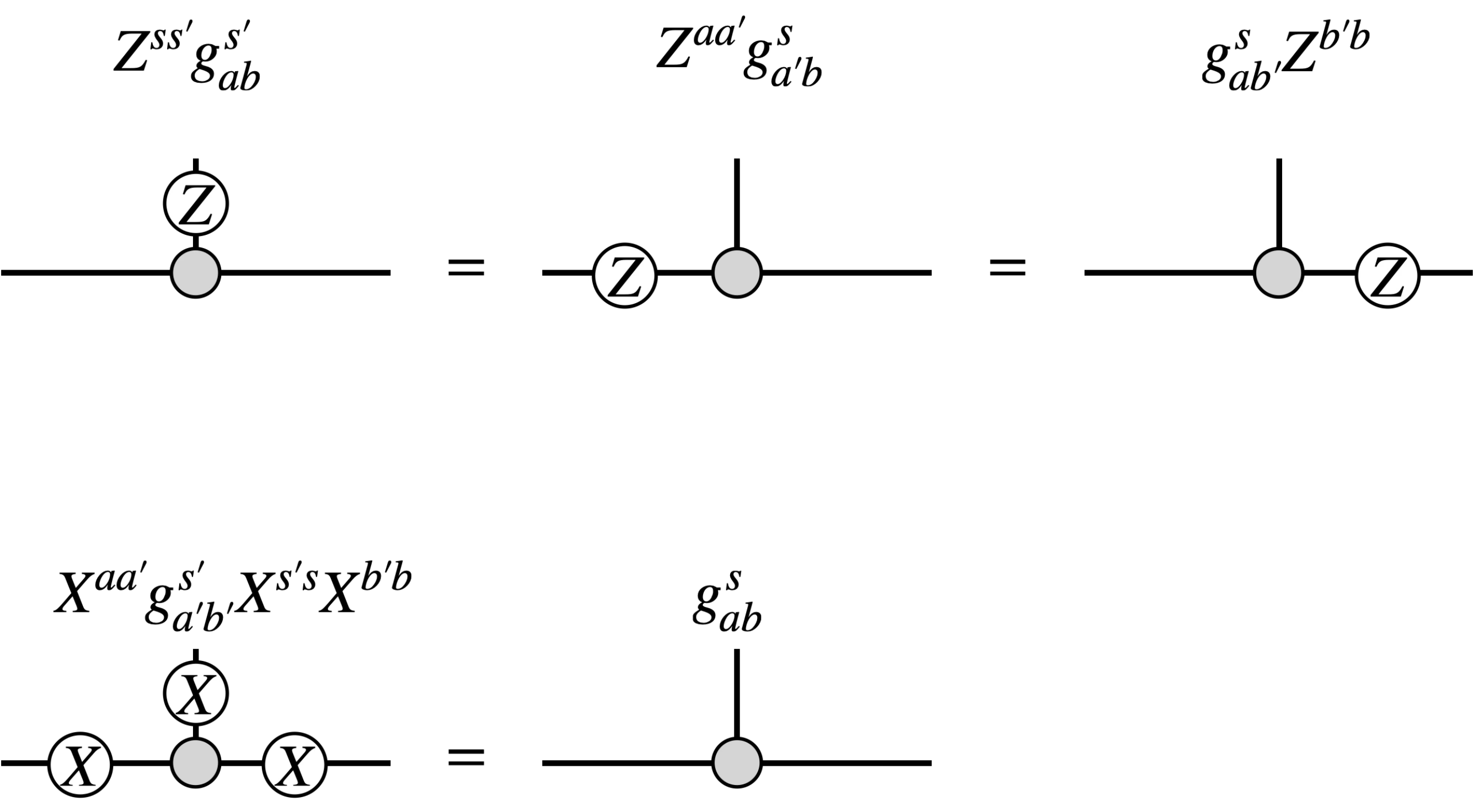}
    \label{eq:tc-inj-g}
\end{align}
and
\begin{align}
    \includegraphics[width=0.76\linewidth]{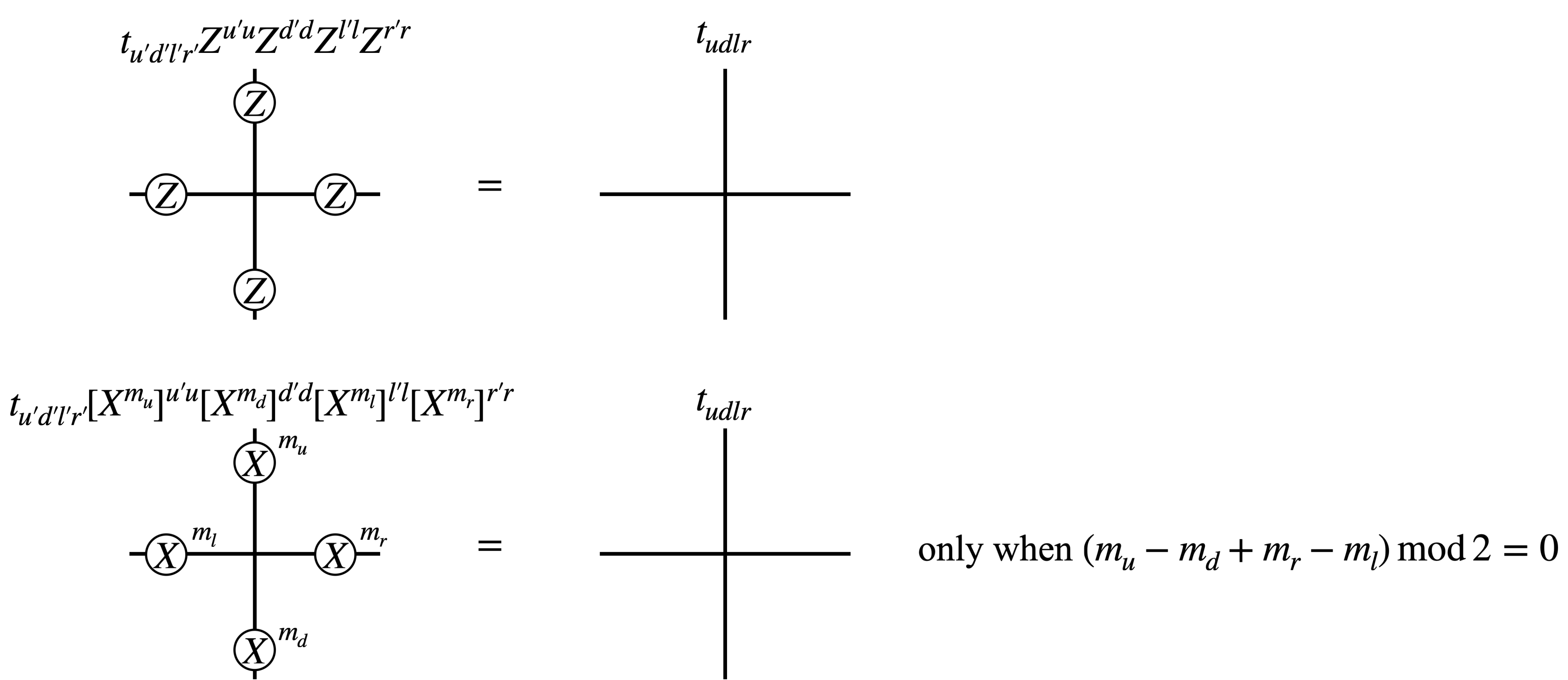}
    \label{eq:tc-inj-t}
\end{align}
Here, \(X\) and \(Z\) are the Pauli matrices.

The local gauge symmetries described in Eqs.\,\eqref{eq:tc-inj-g} and \eqref{eq:tc-inj-t} are inherited by the local tensor \(T_{udlr}^{\tilde{s}}\), resulting in corresponding local gauge symmetries. First, the \(Z\)-injective symmetry of the \(T\) tensor can be expressed as:
\begin{align}
    \includegraphics[width=0.32\linewidth]{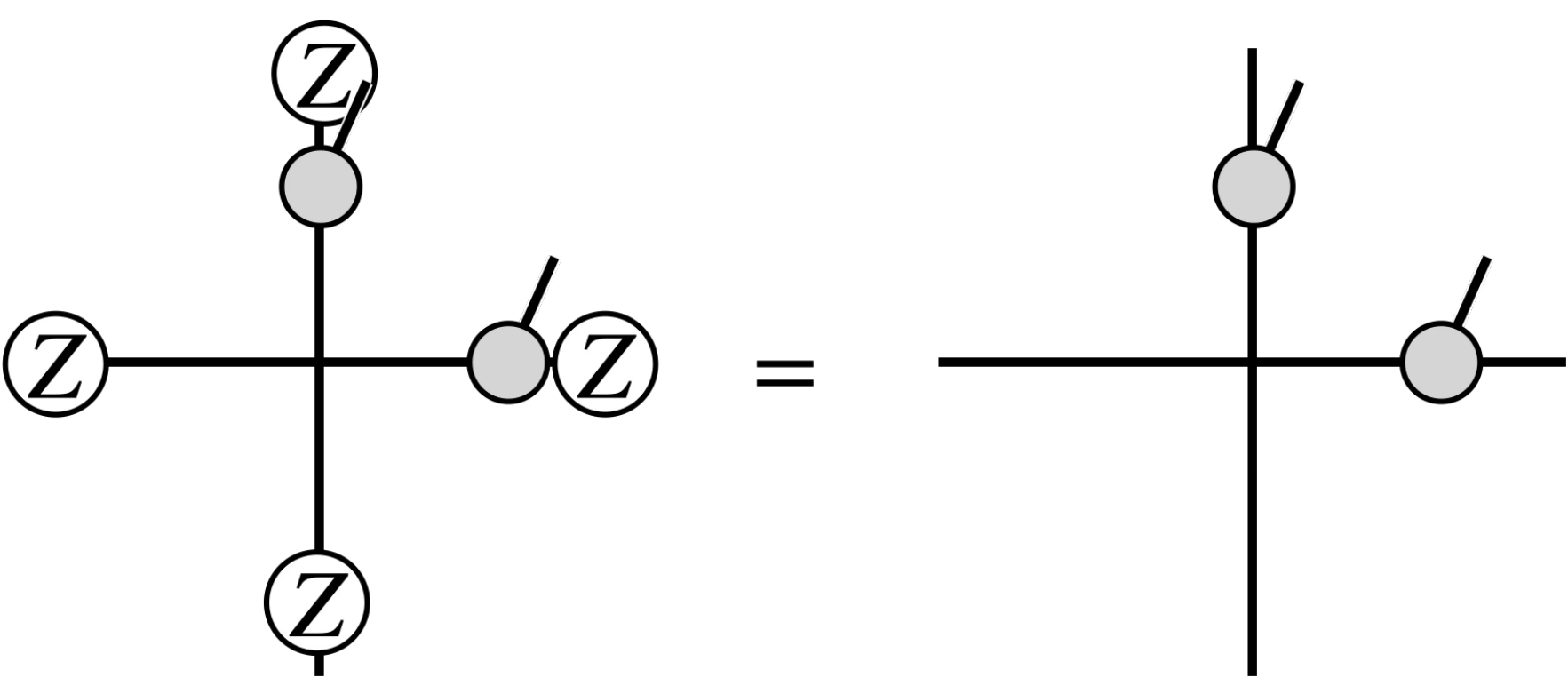}
    \label{eq:tc-T-Z-inj-1}
\end{align}
Additionally, two other useful local gauge symmetries emerge:
\begin{align}
    \includegraphics[width=0.73\linewidth]{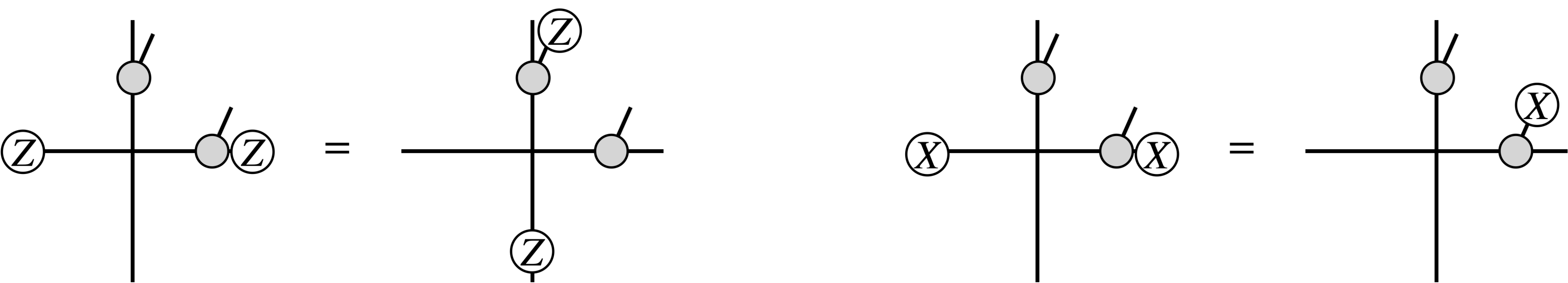}
    \label{eq:tc-T-Z-inj-2}
\end{align}

By contracting the vertical indices of the \(T\) tensors after arranging them in a column, the \({\mathbb R}\) tensor defined in the main text can be obtained. Using the injective symmetry in Eq.\,\eqref{eq:tc-T-Z-inj-1}, it can be shown that the \({\mathbb{C}}\) tensor possesses a one-dimensional {\it global} symmetry at the virtual index level: \( \left(\prod_i Z_i\right){\mathbb{C}} \left(\prod_i Z_i\right) = {\mathbb{C}} \).  
This symmetry can be represented graphically as:
\begin{align}
    \includegraphics[width=0.28\linewidth]{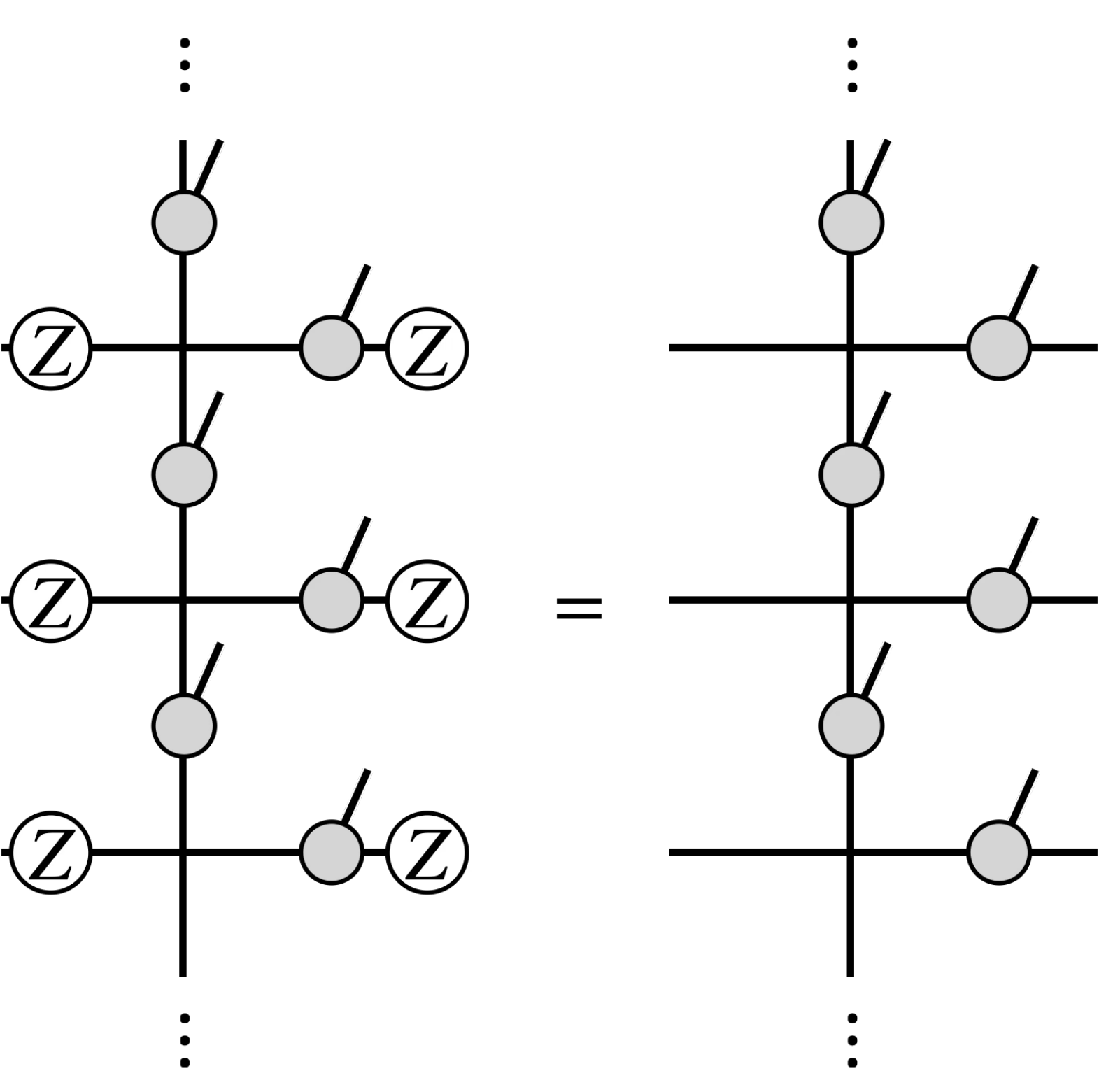}
    \label{eq:tc-R-Z-1}
\end{align}
This global symmetry is directly reflected in the two symmetries \(g_{Z \mathbb{I}}\) and \(g_{ZZ}\) of the one-dimensional column-to-column transfer matrix \({\mathbb H} = {\rm tTr}_{\{s_i\}} \mathbb{C}^{\{s_i\}}{\mathbb{C}^*}^{\{s_i\}} \), such that \(g_{Z\mathbb{I}} {\mathbb H} g_{Z\mathbb{I}} = \mathbb{H} \) and \(g_{ZZ} {\mathbb H} g_{ZZ} = \mathbb{H} \).

Furthermore, Eq.\,\eqref{eq:tc-T-Z-inj-2} shows that the local application of \(X\) or \(Z\) operators on virtual indices is transferred to the physical indices as follows:
\begin{align}
    X_{l_j} \, \mathbb{C} \, X_{r_j} = X_{s^x_j}  \mathbb{C},\quad\quad
    Z_{l_j} \, \mathbb{C} \, Z_{r_j} = Z_{s^1_j} Z_{s^1_{j+1}} \mathbb{C},
    \label{eq:tc-R-XZ-local-1}
\end{align}
or equivalently, in graphical form:
\begin{align}
    \includegraphics[width=0.65\linewidth]{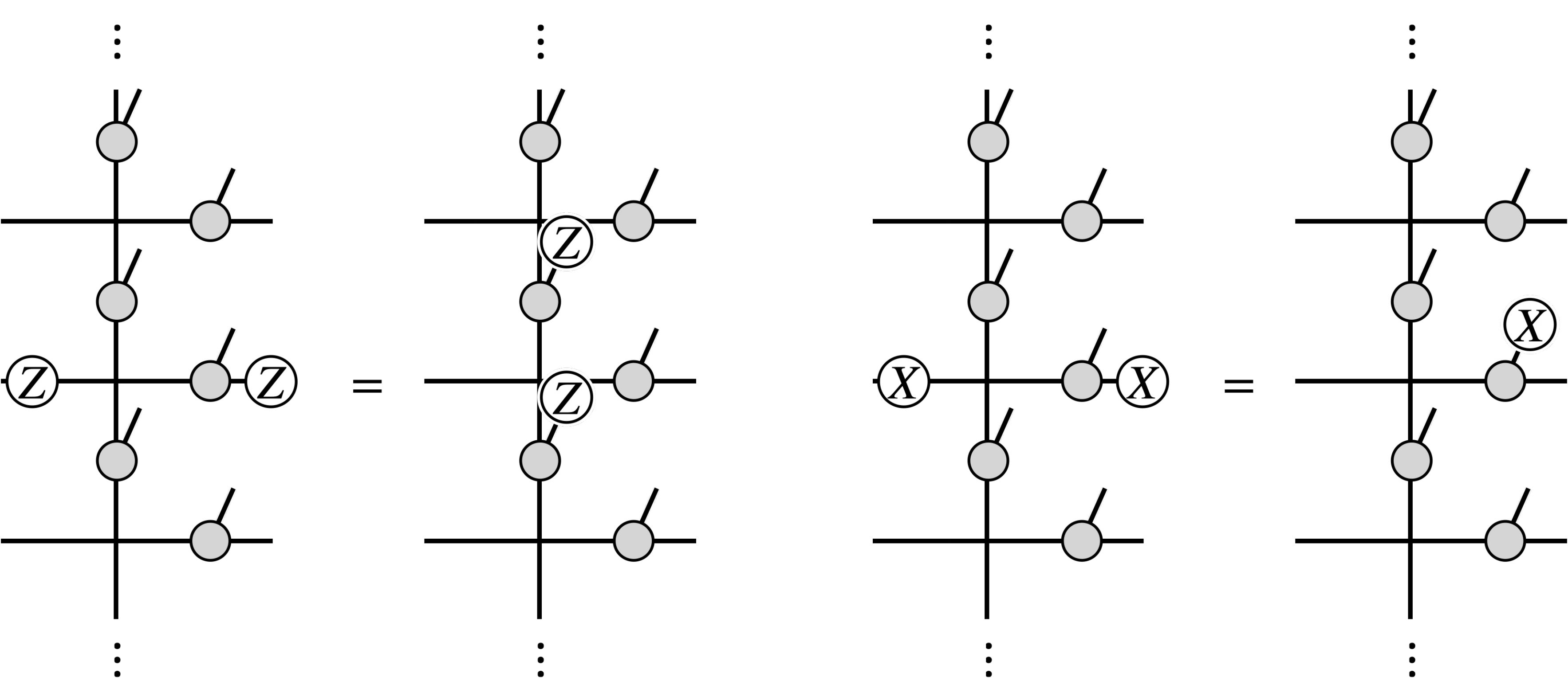}
    \label{eq:tc-R-XZ-local}
\end{align}
Interestingly, during the contraction of physical indices to construct the column-to-column transfer matrix \({\mathbb H}\), the \(X\) and \(Z\) operators cancel out. This cancellation gives rise to local symmetries in the transfer matrix, defined as \(u_i = Z_i \otimes \bar{Z}_i\) and \(v_i = X_i \otimes \bar{X}_i\). Here, operators without a bar act on the ket layer of the transfer matrix, while operators with a bar act on the bra layer.

\subsection{Deformed Toric Code Wavefunction}
By applying the filtering operator \(D_i(\beta_x, \beta_z) = I_i + \beta_x X_i + \beta_z Z_i\) to each physical degree of freedom, the filtered TC state can be obtained as: \(|{\rm fTC(\beta_x, \beta_z) \rangle = \prod_i D_i(\beta_x, \beta_z) |{\rm TC}\rangle }\).

The filtering operator locally modifies the \(\mathbb{Z}_2\) physical basis as follows:
\begin{align}
|0\rangle \rightarrow |\tilde{0}\rangle  & \propto \begin{pmatrix} 1+\beta_z \\ \beta_x \end{pmatrix}
\nn
|1\rangle \rightarrow |\tilde{1}\rangle  & \propto \begin{pmatrix} \beta_x \\ 1-\beta_z \end{pmatrix}
\end{align}
After filtering, the local states are no longer orthonormal. To simplify the analysis, we set \(\langle \tilde{0} |\tilde{0}\rangle = 1\), such that the overlaps between the local states are given by:
\begin{align}
\langle \tilde{0} |\tilde{1} \rangle = \frac{\beta_x^2 + (\beta_z -1)^2}{\beta_x^2 + (\beta_z+1)^2}, 
\quad 
\langle \tilde{1}|\tilde{1}\rangle = \frac{2 \beta_x}{\beta_x^2 + (\beta_z+1)^2}.
\end{align}

It is important to note that the filtering operator modifies only the smaller tensor \(g\), leaving \(t\) unchanged. Consequently, the \(Z\)-injectivity property in Eq.\,\eqref{eq:tc-T-Z-inj-1} remains valid across the entire range of filtering parameters. However, the \(X\) local gauge symmetry in Eq.\,\eqref{eq:tc-T-Z-inj-2} holds only when \(\beta_z = 0\), and the \(Z\) local gauge symmetry in Eq.\,\eqref{eq:tc-T-Z-inj-2} holds only when \(\beta_x = 0\). 

As a result, the column-to-column transfer matrix \(\mathbb{H}\) exhibits the local symmetry \(u_i = X_i \otimes \bar{X}_i\) only for \(\beta_z = 0\), and the local symmetry \(v_i = Z_i \otimes \bar{Z}_i\) only for \(\beta_x = 0\).

\subsection{Emergent local and global symmetries of transfer matrix}
At specific points or along certain lines, the transfer matrix \(\mathbb{H}\) exhibits additional local or global symmetries, which are valuable for analyzing the phase diagram of the filtered toric code.

Two special points give rise to additional local symmetries. The first occurs at the point \([\beta_z, \beta_x] = [0,1]\), where \(|\tilde{0}\rangle = |\tilde{1}\rangle\), making the two local physical states indistinguishable. This leads to the following:
\begin{align}
    \includegraphics[width=0.65\linewidth]{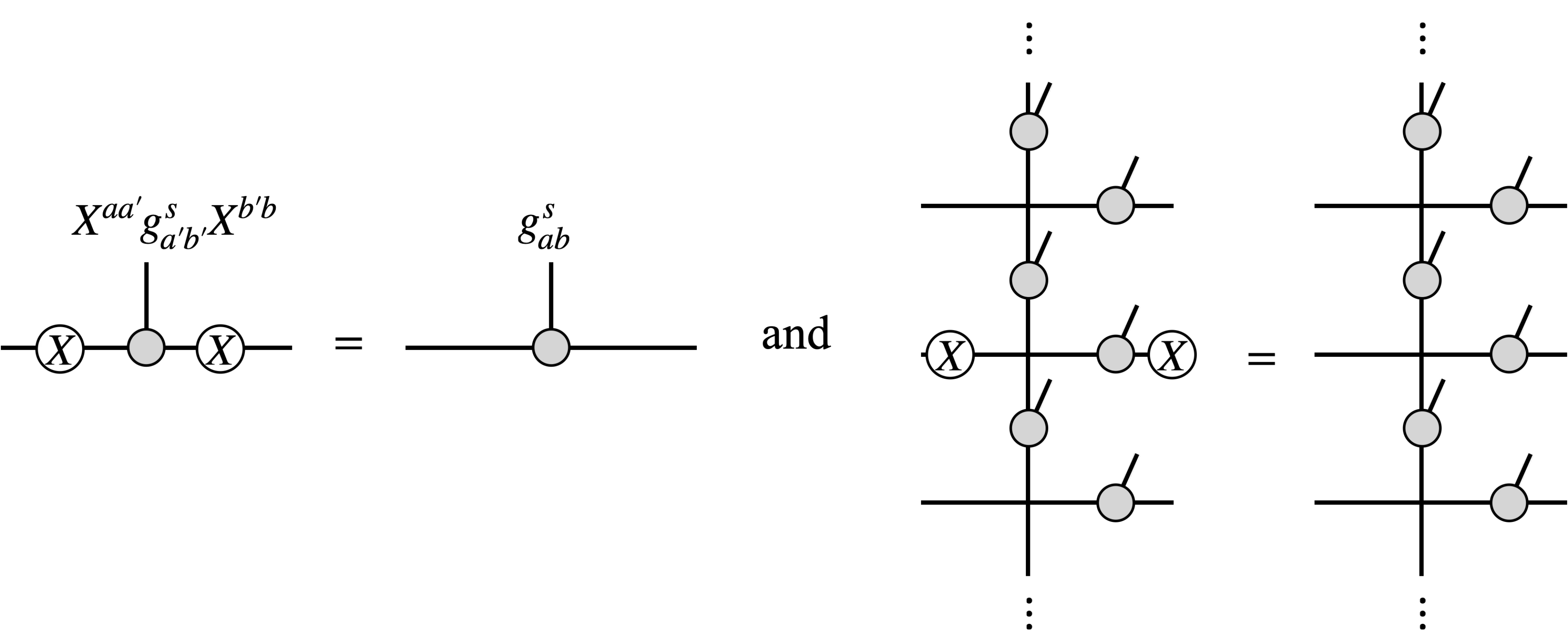}
    \label{eq:tc-beta_x-0}
\end{align}
As a result, the transfer matrix attains a local symmetry \(m_i^1 = X_i \otimes \bar{\mathbb{I}}_i\) at \([\beta_z, \beta_x] = [0,1]\).

The second special point occurs at \([\beta_z, \beta_x] = [1,0]\), where \(|\tilde{1}\rangle = 0\). In this case, we get:
\begin{align}
    \includegraphics[width=0.75\linewidth]{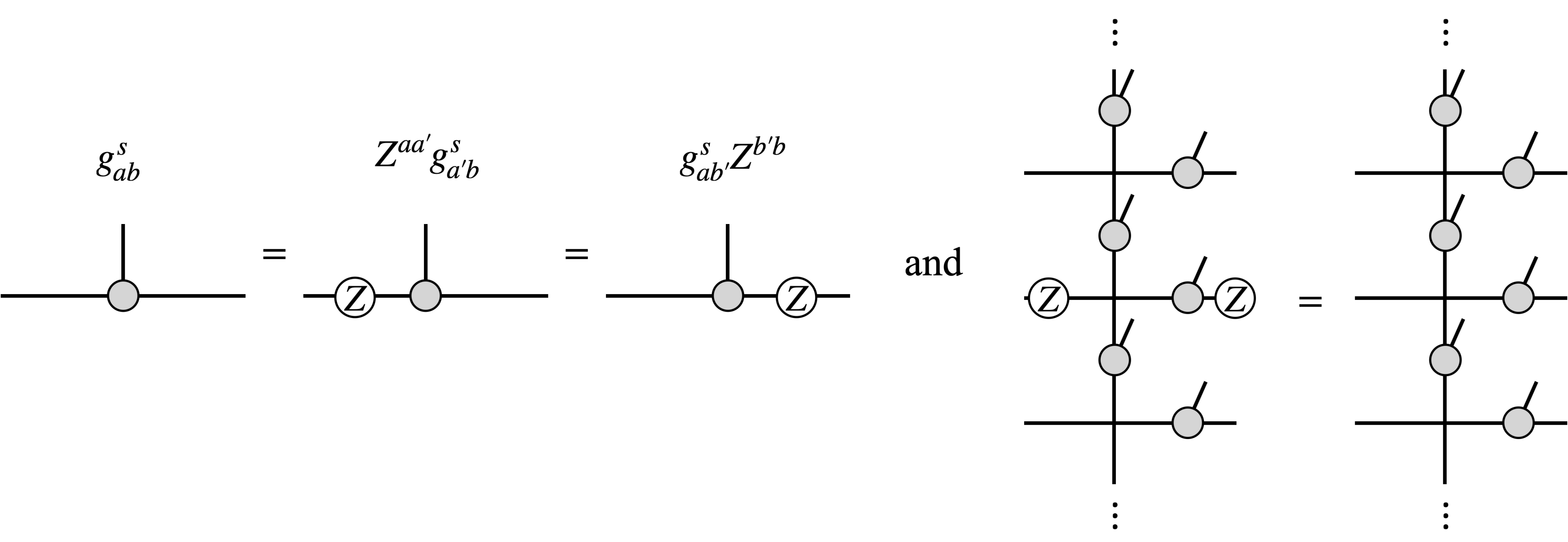}
    \label{eq:tc-beta_z-0}
\end{align}
Here, the transfer matrix \(\mathbb{H}\) achieves a local symmetry \(u_i^1 = Z_i \otimes \bar{\mathbb{I}}\) at \([\beta_z, \beta_x] = [1,0]\).

In contrast, when \(\langle \tilde{0} |\tilde{1} \rangle = \langle \tilde{1} | \tilde{1}\rangle\), the transfer matrix \(\mathbb{H}\) gains an additional global symmetry, alongside the existing symmetries \(g_{ZI}\) and \(g_{ZZ}\). This condition is satisfied along the line:
\begin{align}
(\beta_x -1)^2 + (\beta_z-1)^2 = 1.
\label{eq:tc-potts-line}
\end{align}
It is important to note that the line described by Eq.\,\eqref{eq:tc-potts-line} includes the points \([\beta_z, \beta_x] = [1,0]\) and \([\beta_z, \beta_x] = [0,1]\).

To verify that the transfer matrix \(\mathbb{H}\) acquires an additional global symmetry along the line, it is helpful to express the transfer matrix as a tensor product of local tensors: 
\[
\mathbb{H}^{\{\bar{l}_i\}, \{\bar{r}_i\}}_{\{l_i\}, \{r_i\}} = {\rm tTr}_{u_i, d_i, \bar{u}_i, \bar{d}_i} \mathbb{E}_{u_i l_i d_i r_i}^{\bar{u}_i \bar{l}_i \bar{d}_i \bar{r}_i}.
\]
The local tensor \(\mathbb{E}_{u_i l_i d_i r_i}^{\bar{u}_i \bar{l}_i \bar{d}_i \bar{r}_i}\) is given by:
\[
\mathbb{E}_{u_i l_i d_i r_i}^{\bar{u}_i \bar{l}_i \bar{d}_i \bar{r}_i} = \sum_{s_i} T^{s_i}_{u_i l_i d_i r_i} (T^*)^{s_i}_{\bar{u}_i \bar{l}_i \bar{d}_i \bar{r}_i},
\]
or graphically represented as:
\begin{align}
    \includegraphics[width=0.42\linewidth]{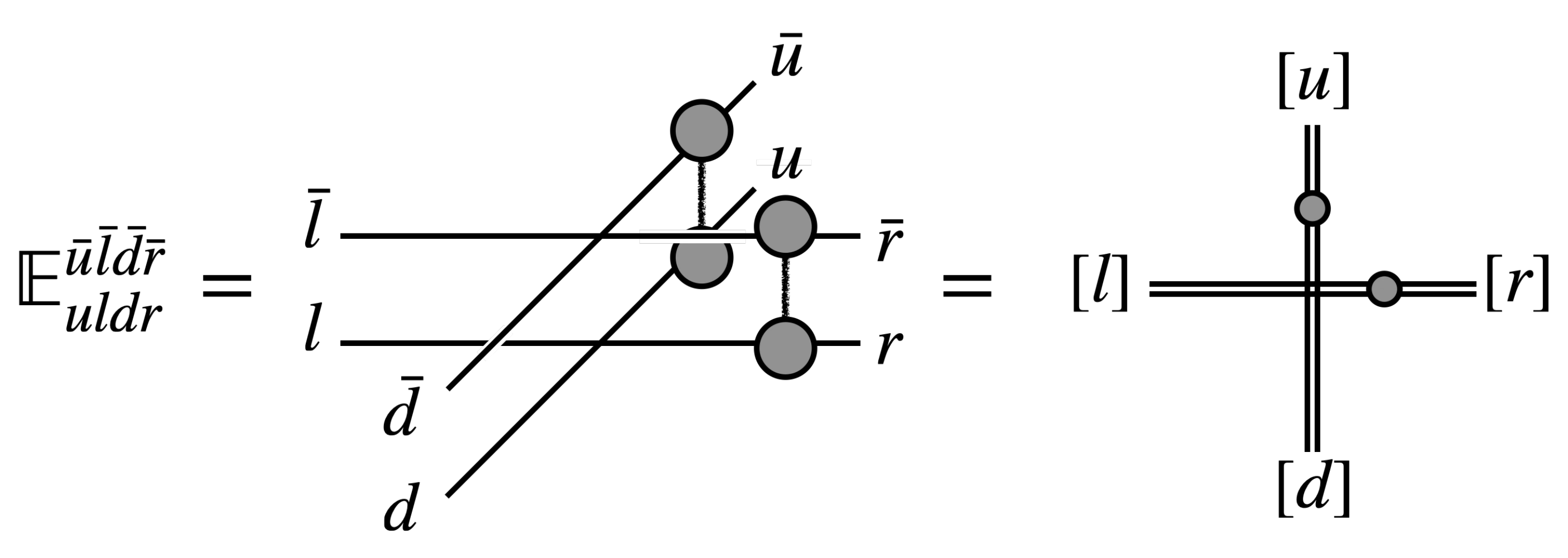}
    \label{eq:tc-E}
\end{align}
In this expression, the indices with and without bars are grouped, i.e., \([a] = [a, \bar{a}]\). Each of these grouped indices can take four possible values: \([0,\bar{0}],\, [0,\bar{1}], \, [1,\bar{0}]\), and \([1,\bar{1}]\).

Within this four-dimensional index space, one can define an operator \(P\) with the following matrix representation:
\begin{align}
P = 
\begin{pmatrix}
1 & 0 & 0 & 0 \\
0 & 0 & 0 & 0 \\
0 & 0 & 0 & 0 \\
0 & 0 & 0 & 0
\end{pmatrix}.
\end{align}
This operator permutes the three basis states of the four-dimensional local index space in the sequence: \([0,\bar{1}]\rightarrow[1,\bar{0}]\rightarrow[1,\bar{1}]\rightarrow[0,\bar{1}]\).

Upon closer inspection, it becomes clear that the local tensor \(\mathbb{E}\) exhibits injective symmetry under \(P\), i.e.,
\begin{align}
    \includegraphics[width=0.60\linewidth]{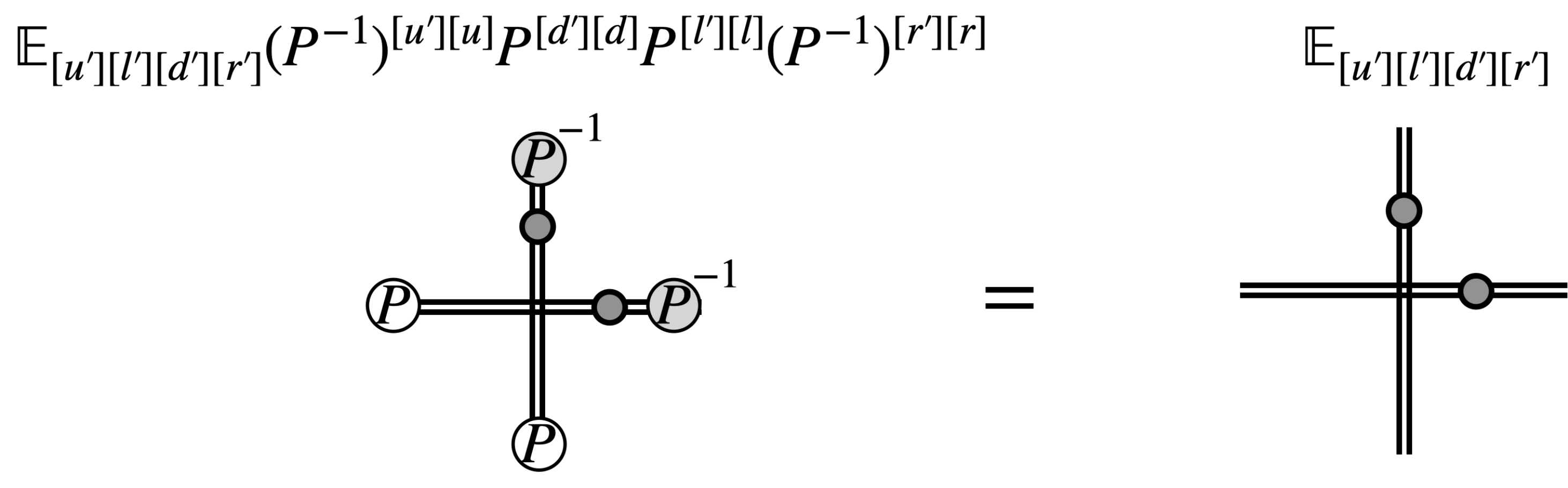}
    \label{eq:tc-E-inj}
\end{align}
From this injectivity, we deduce that the transfer matrix \(\mathbb{H}\) is symmetric under the global symmetry operator \(g_P = \prod_i P_i\), meaning that \(g_P^{-1} \mathbb{H} g_P = \mathbb{H}\).

By expressing the local order parameter operators as
\begin{align}
X_i \otimes \bar{I}_i = 
\begin{pmatrix}
0 & 1 & 0 & 0 \\
1 & 0 & 0 & 0 \\
0 & 0 & 0 & 1 \\
0 & 0 & 1 & 0
\end{pmatrix}, \quad 
I_i \otimes \bar{X}_i = 
\begin{pmatrix}
0 & 0 & 1 & 0 \\
0 & 0 & 0 & 1 \\
1 & 0 & 0 & 0 \\
0 & 1 & 0 & 0
\end{pmatrix},
\end{align}
it is straightforward to demonstrate the following relations:
\begin{align}
P \left(X_i \otimes \bar{I}_i\right) P^{-1} & = I_i \otimes \bar{X}_i, \nn
P \left(I_i \otimes \bar{X}_i\right) P^{-1} & = X_i \otimes \bar{X}_i, \nn
P \left(X_i \otimes \bar{X}_i\right) P^{-1} & = I_i \otimes \bar{X}_i.
\end{align}
In other words, along the \(g_P\)-symmetric line, the local order parameters \(I_i \otimes \bar{X}_i\) and \(X_i \otimes \bar{X}_i\) become equivalent. This implies that the corresponding global symmetries \(g_{ZZ}\) and \(g_{Z\mathbb{I}}\) are either both simultaneously broken or both preserved.

\section{Kitaev Honeycomb Loop Gas Operator}

%
\subsection{Gauge Symmetry and Physical-Virtual Relation}

The loop gas\,(LG) operator for the Kitaev honeycomb model is efficiently represented by a tensor product form\,\cite{hy19} with a bond dimension $D=2$: $ \hat{Q}_{\rm LG}(\delta) = {\rm tTr} \prod_{\alpha} Q_{i_{\alpha} j_{\alpha} k_{\alpha}}^{ss'}(\delta)|s\rangle \langle s'|$ with a building block tensor
\begin{align}
    Q_{i j k}^{ss'}(\delta) = \tau_{i j k }(\delta) [X^{1-i} Y^{1-j} Z^{1-k}]_{ss'},
	\label{eq:q_operator}
\end{align}
where $X, Y, Z$ are the Pauli matrices, the virtual indices $i,j,k=0,1$, and non-zero elements of $\tau$-tensor are 
\begin{align}
    \tau_{000} = -i, \quad  \tau_{011}=\tau_{101}=\tau_{110}=\delta.
\end{align}
It is important to note that the $\tau$-tensor contains non-zero elements only when the count of index-1 is even. Consequently, regardless of the variational parameter $\delta$, the $Q$-tensor remains invariant under the action of $Z^{\otimes 3}$ on its virtual indices, i.e., $Z_{ii'} Z_{jj'} Z_{kk'} Q_{i'j'k'}^{ss'}(\delta) = Q_{ijk}^{ss'}(\delta)$ or graphically
\begin{align}
    \includegraphics[width=0.7\linewidth]{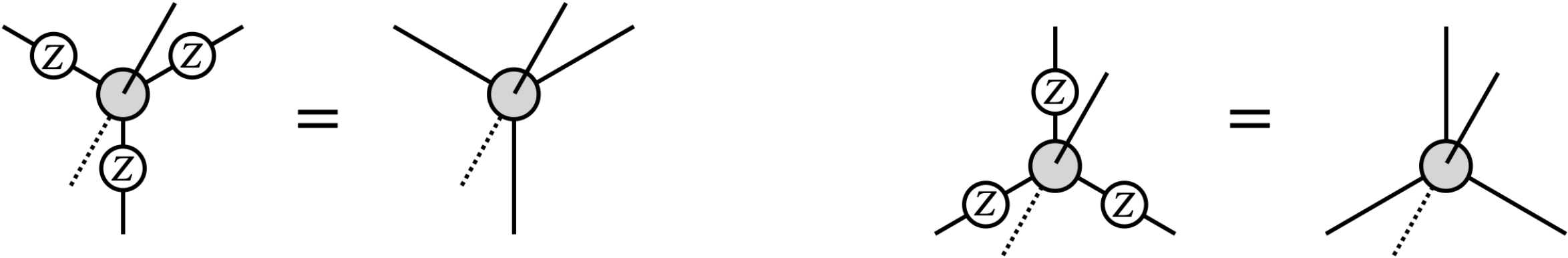}
    \label{eq:q_gauge_sym}
\end{align}
Furthermore, at \(\delta=1\), the \(Q\)-tensor satisfies remarkable properties, referred to as physical-virtual relations, wherein the action of Pauli matrices on the physical index can be equivalently represented by a unitary transformation on the virtual indices.:
\begin{align}
    X_{s\tilde{s}} \,Q_{ijk}^{\tilde{s} s'} (1)
    = v_{jj'} v_{kk'}^* \,Q_{ij'k'}^{ss'}(1)
    = v_{jj'}^* v_{kk'} Z_{ii'} \, Q_{i'j'k'}^{ss'}(1), \nonumber \\ 
	Y_{s\tilde{s}} \, Q_{ijk}^{\tilde{s} s'} (1)
    = v_{kk'} v_{ii'}^* \, Q_{i'jk'}^{ss'}(1)
    = v_{kk'}^* v_{ii'} Z_{jj'} \, Q_{i'j'k'}^{ss'}(1), \nonumber\\
	Z_{s\tilde{s}} \, Q_{ijk}^{\tilde{s} s'} (1)
    = v_{ii'} v_{jj'}^* \, Q_{i'j'k}^{ss'}(1)
    = v_{ii'}^* v_{jj'} Z_{kk'} \, Q_{i'j'k'}^{ss'}(1),
\end{align}
or graphically
\begin{align}
    \includegraphics[width=0.4\linewidth]{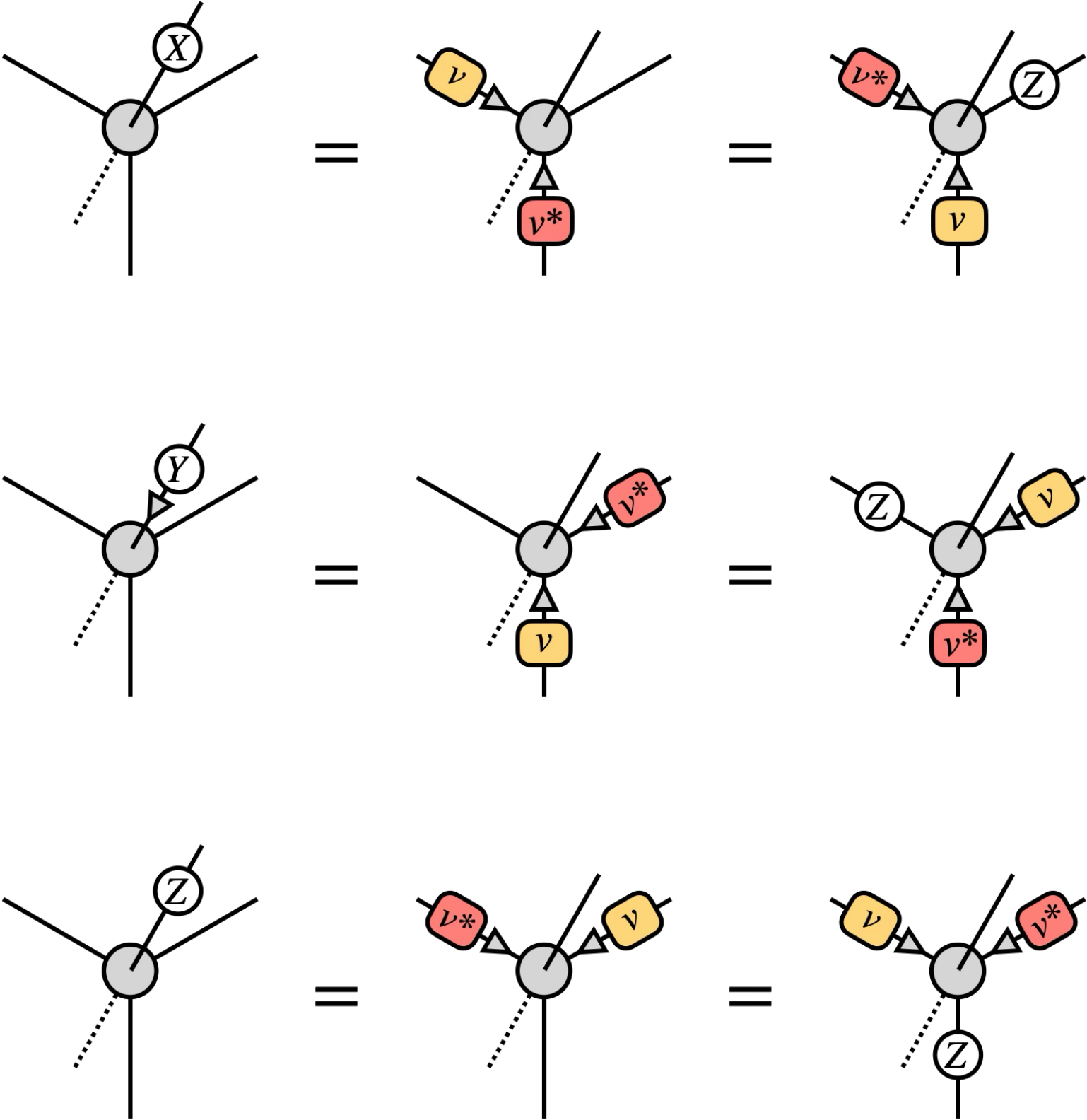}
    \label{eq:q_phys_virt},
\end{align}
where the unitary matrix 
\begin{align}
    v = \begin{pmatrix}
        0 & i \\ 1 & 0
    \end{pmatrix}
    =\frac{1}{2} (X - Y) + \frac{i}{2} (X - Y),
\end{align}
Here, the direction indicated in the graphical representation corresponds to the direction of matrix multiplication. Specifically, if the direction points outward\,(inward) from the matrix, then its column\,(row) index is contracted. For non-symmetric matrices ($M^t \neq M$), the direction of the contraction must be carefully managed to ensure consistency. We omit the direction for symmetric matrices as it becomes redundant.
Remarkably, using the first equalities in Eq.\,\eqref{eq:q_phys_virt}, it can be shown that the LG operator functions as a projector, mapping any state onto the flux-free sector, expressed as $\hat{W}_p \, \hat{Q}_{\rm LG}(1) = \hat{Q}_{\rm LG}(1)\, \hat{W}_p = \hat{Q}_{\rm LG}(1)$\,\cite{hy19}. On the other hand, the set of second equalities, which have not been previously recognized, play a pivotal role in the central results of this work alongside the set of first equalities.

\subsection{Global and Local symmetries of the Transfer Matrix}

To discuss the global and local symmetries of the transfer matrix, we begin by considering the double-layer of the LG operator, \(\hat{Q}_{\rm LG} \hat{Q}_{\rm LG}^\dagger\), and its column-to-column operator, as illustrated below:
\begin{align}
    \includegraphics[width=0.3\linewidth]{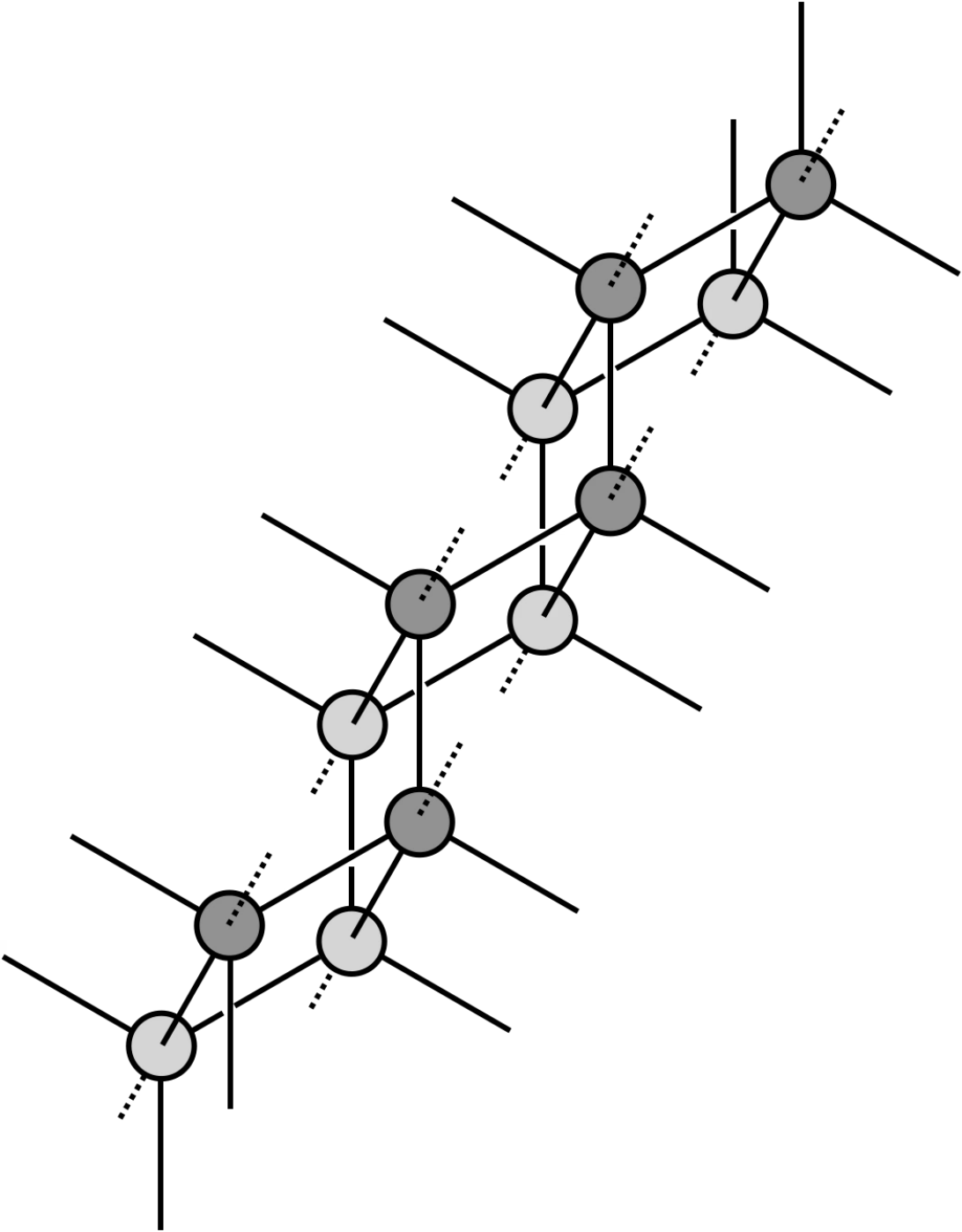}.
\end{align}
Using the gauge symmetry of the \(Q\)-tensor described in Eq.\,\eqref{eq:q_gauge_sym}, it is straightforward to show that a global unitary transformation \(g_{Z\mathbb{I}} \equiv (Z\otimes \mathbb{I}) ^{\otimes L}\), acting on the virtual indices in the double-layer operator, leaves it invariant, as illustrated below:
\begin{align}
    \includegraphics[width=0.9\linewidth]{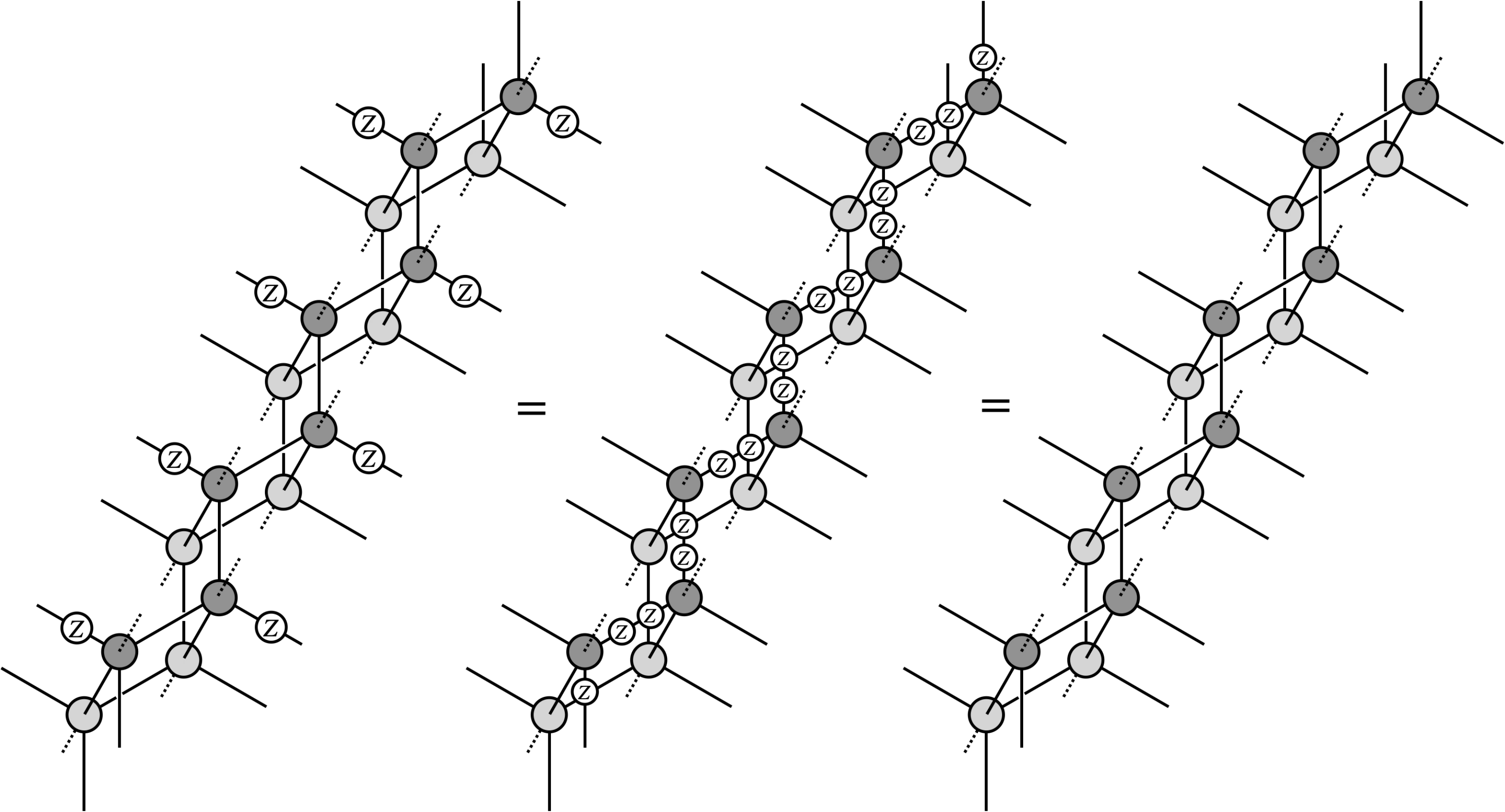}.
\end{align}
Similarly, another global operation, \( g_{ZZ} \equiv (Z \otimes Z)^{\otimes L} \), also leaves the double-layer operator invariant. Consequently, the transfer matrix \(\mathbb{H}\) of a quantum state \( |\psi\rangle = \hat{Q}_{\rm LG} |\varphi\rangle \) possesses the global \(\mathbb{Z}_2 \times \mathbb{Z}_2\) symmetry generated by \(g_{Z\mathbb{I}}\) and \(g_{ZZ}\), provided that the state \( |\varphi\rangle \) is trivial. More specifically, \( |\varphi\rangle \) can be represented by a PEPS in which the tensor does not exhibit gauge symmetry, regardless of whether it is a product state or an entangled state.

Now, we show that the column-to-column operator is invariant under a local unitary transformation ${\bm v}_i \equiv (v_{i-1}^* \otimes v_{i-1}^t) (v_{i} \otimes v_{i}^\dagger) $ as illustrated below;
\begin{align}
    \includegraphics[width=0.5\linewidth]{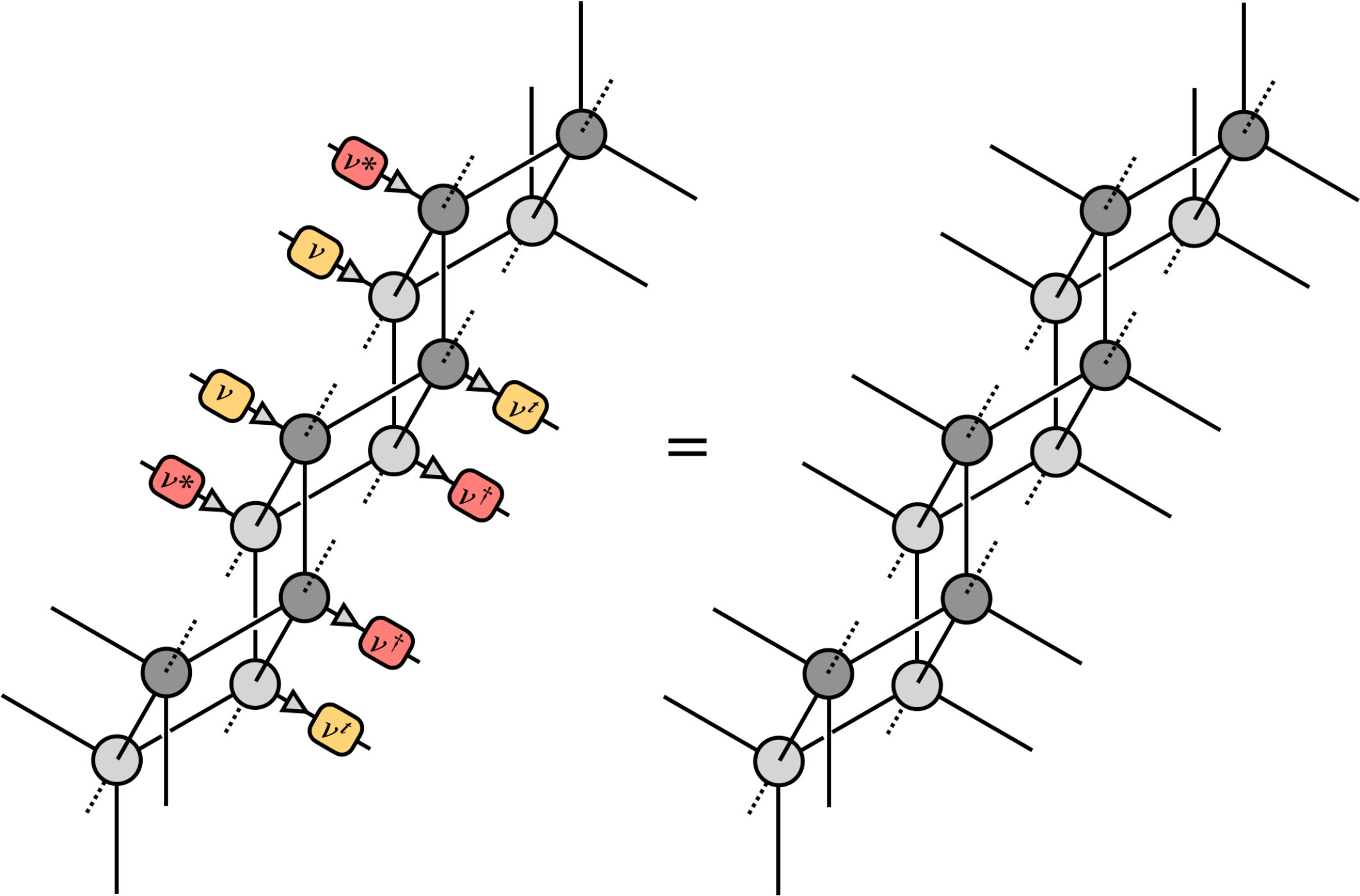}.
\end{align}
The local unitary transformation \(v_{i-1}^* \, v_i\) acting on the virtual indices of the LG operator is equivalent to the action of a Pauli matrix on the corresponding physical indices, as illustrated below:
\begin{align}
    \includegraphics[width=0.5\linewidth]{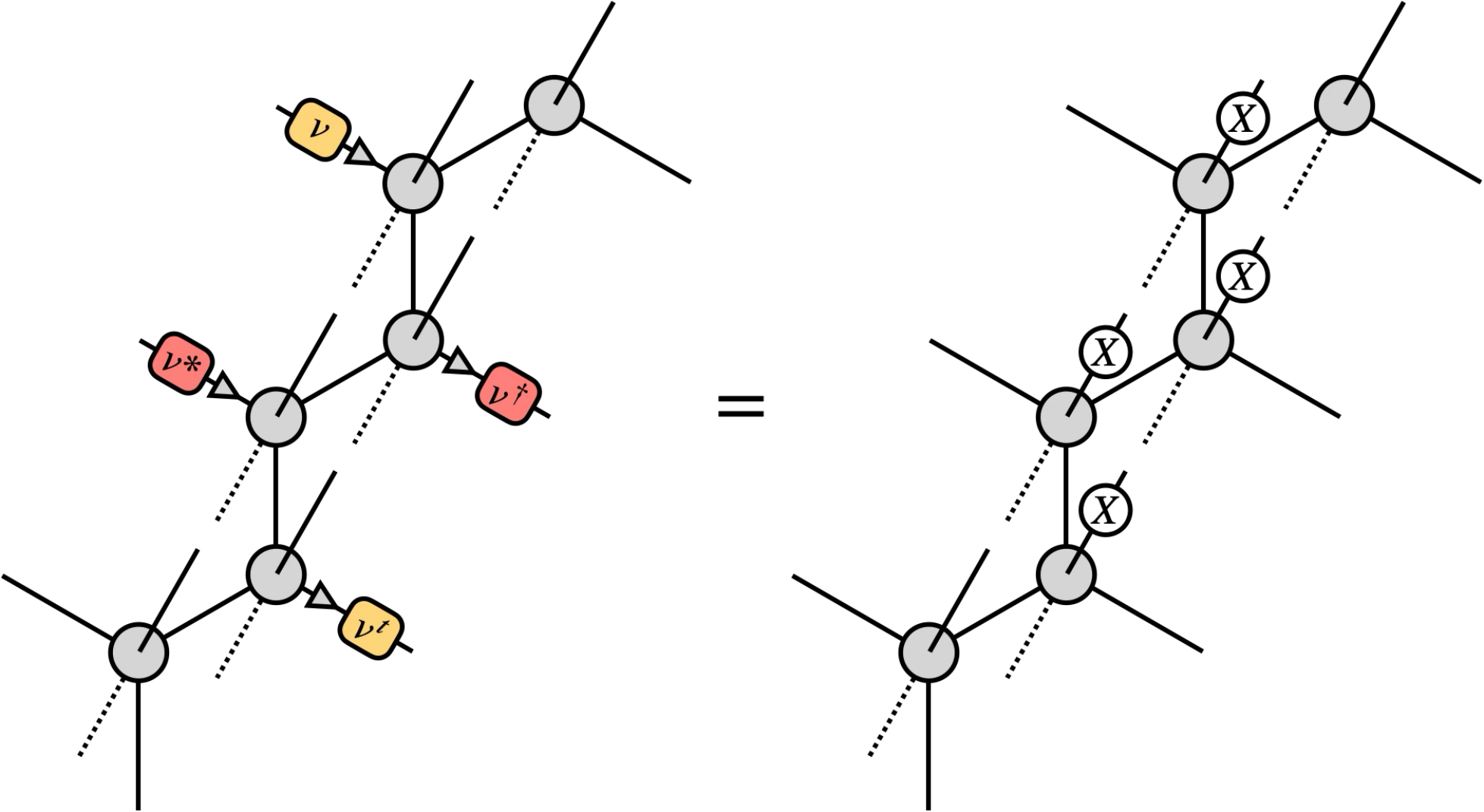}
    \label{eq:lg-unitary-2site},
\end{align}
where we use Eq.\,\eqref{eq:q_phys_virt} to derive the following two identities:
\begin{align}
    \includegraphics[width=0.9\linewidth]{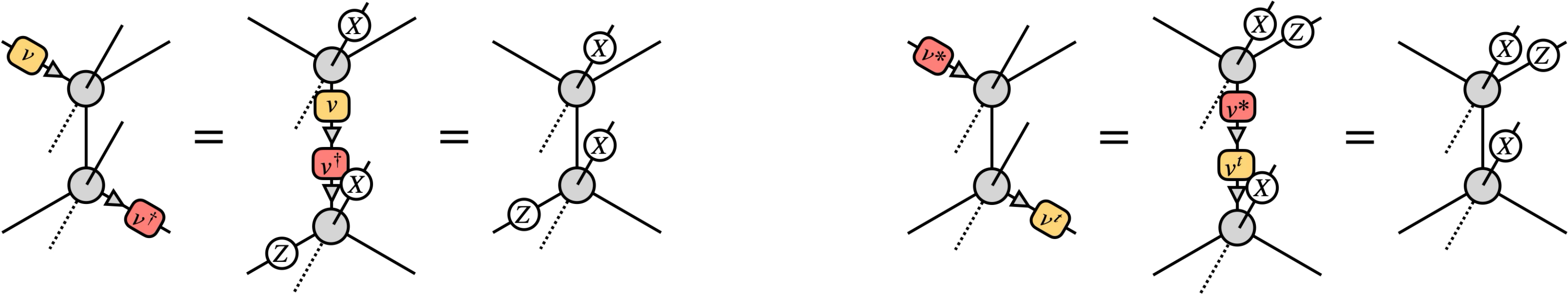}.
\end{align}
Therefore, the double-layer operator is symmetric under the action of the \({\bm v}_i\) operator, as the Pauli \(X\) matrices acting on the physical indices cancel out during the contraction of physical indices.

\subsection{Variational Ansatz}

The variational ansatz we consider is as follows:
\begin{align}
    |{\rm LG}(\theta, \delta)\rangle = \hat{Q}_{\rm LG}(\delta) |\theta\rangle^{\otimes N},    
\end{align}
where the product state $|\theta\rangle^{\otimes N} \equiv \prod_{n=1}^N |\theta\rangle_n$ with $|\theta\rangle$ being the lowest eigenstate of the Hamiltonian
\begin{align}
    h(\theta) \equiv - \cos\theta Z -\sin\theta (X+Y).
\end{align}
By contracting tensors in a unit-cell, 
\begin{align}
    T_{iji'j'}^{s_1 s_2}(\delta) = \sum_{s_1', s_2'}\sum_k Q_{ijk}^{s_1 s_1'}(\delta) Q_{i'j'k}^{s_2 s_2'} (\delta)\langle s_1' | \theta\rangle  \langle s_2' | \theta\rangle,
\end{align}
or graphically
\begin{align}
    \includegraphics[width=0.3\linewidth]{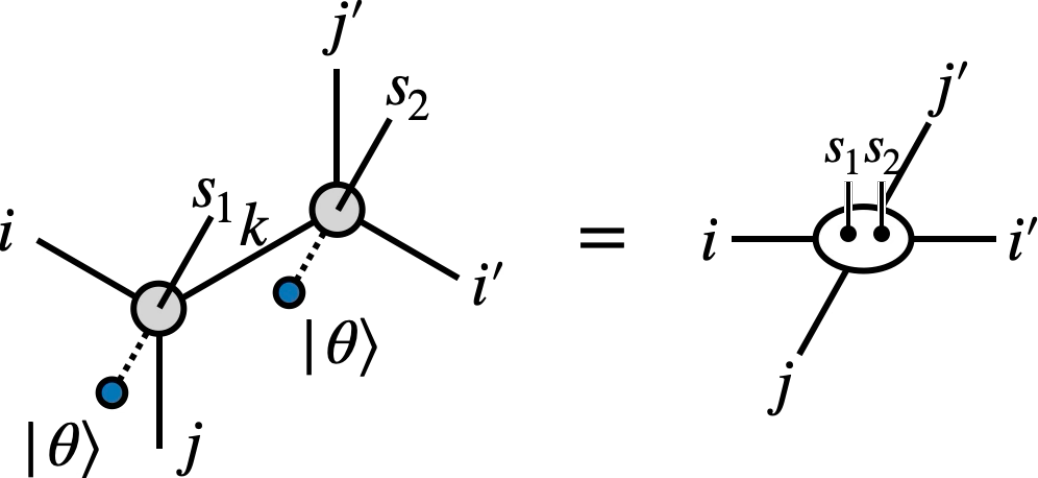},
\end{align}
the network transforms into a square lattice. At \((\theta, \delta) = (0,1)\), the ansatz becomes the exact ground state of the Kitaev honeycomb model in the strong anisotropic limit (\(J_z \gg J_x, J_y\)), where the model maps to Wen's plaquette model, \(H_{\rm Wen} = - \sum_p Y_{1_p} Z_{2_p}Y_{3_p} Z_{4_p}\). Conversely, at \((\theta, \delta) = (\frac{\pi}{4},1)\), the ansatz corresponds to the deconfined quantum critical state characterized by the \(2d\) Ising CFT, which separates two gapped phases for \(\delta < 1\) and \(\delta > 1\) at \(\theta = \frac{\pi}{4}\).

\section{$3d$ Toric code and X-Cube states}
\subsection{PEPS}

The $3d$ toric code\,(3TC) and X-cube\,(XC) states can be efficiently represented by three-dimensional PEPS on the cubic lattice with the following building block tensors:
\begin{align}
    \includegraphics[width=0.7\linewidth]{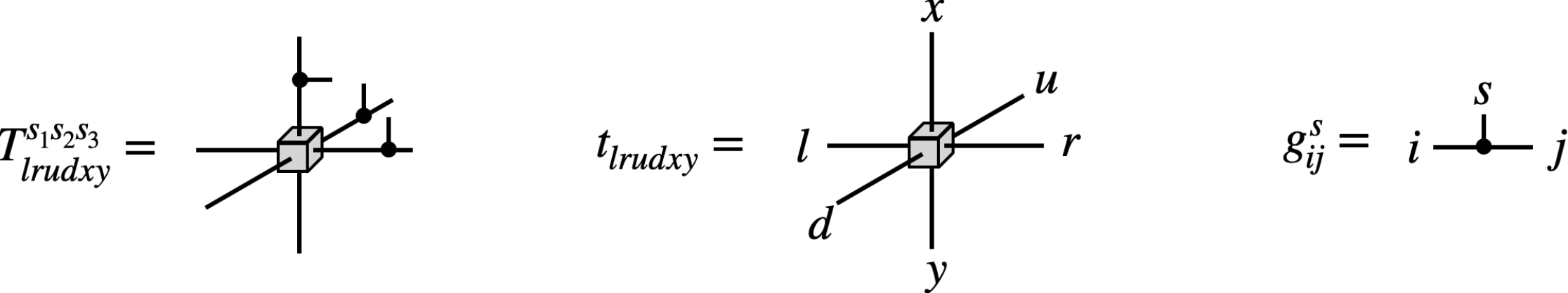},
\end{align}
where $g$-tensor is the same as the one defined in the $2d$ toric code state in Eq.\,\eqref{eq:g_tensor} for both 3TC and XC states. On the other hand, the definition of $t$-tensor for each state is the following\cite{huan18}:
\begin{align}
    t_{lrudxy}^{\rm 3TC} =  
    \begin{cases}
        1, \quad {\rm if} \quad l+r+u+d+x+y = 0 \,\, {\rm mod}\,\, 2 \\
        0, \quad {\rm otherwise}
    \end{cases},
    \quad
    t_{lrudxy}^{\rm XC} =  
    \begin{cases}
        1, \quad {\rm if} \quad 
        \begin{cases}
            l+r+u+d = 0 \,\, {\rm mod}\,\, 2 \\
            l+r+x+y = 0 \,\, {\rm mod}\,\, 2 \\
            u+d+x+y = 0 \,\, {\rm mod}\,\, 2 \\
        \end{cases}
        \\
        0, \quad {\rm otherwise}
    \end{cases}.
\end{align}
Using the definition above, one can easily show that the $t$-tensor for the 3TC state is invariant under the action of the Pauli $Z$ on all virtual indices and the action of $X$ on a pair of virtual indices, as graphically illustrated below:
\begin{align}
    \includegraphics[width=0.5\linewidth]{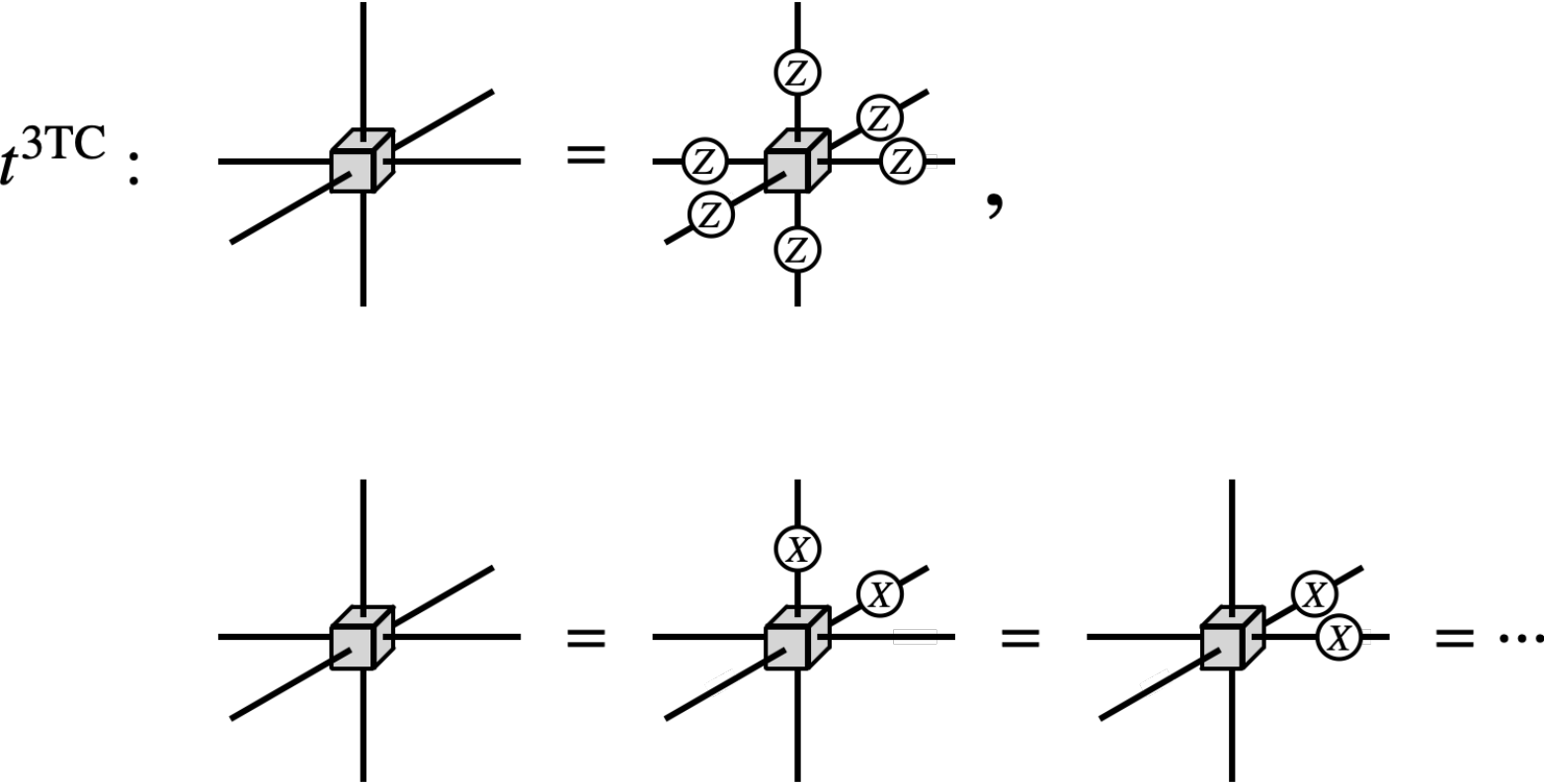}
    \label{eq:3tc_t_tensor_symmetry}
\end{align}
On the other hand, the one for the XC state is invariant under the following transformations:
\begin{align}
    \includegraphics[width=0.8\linewidth]{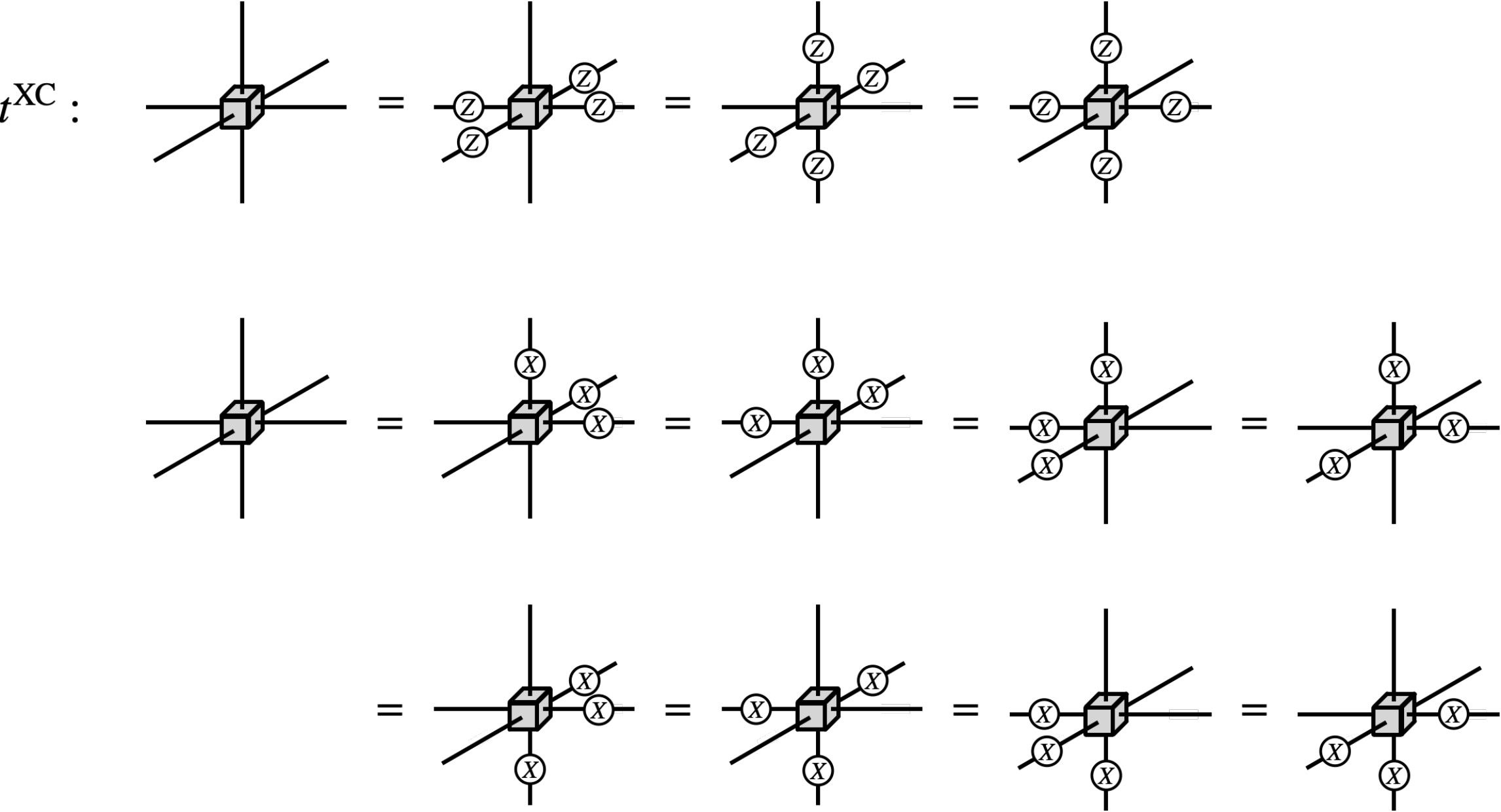}
    \label{eq:xc_t_tensor_symmetry}
\end{align}
\subsection{Global, local symmetries and idempotence of transfer matrices }

Now, we construct the transfer matrices\,(TM) of the norm of the 3TC and XC states, which can be represented by a two-dimensional projected entangled-pair operator (PEPO) consisting of a building block tensor illustrated as:
\begin{align}
    \includegraphics[width=0.15\linewidth]{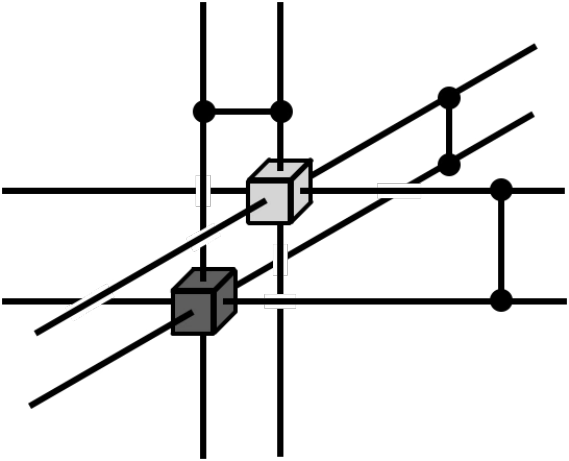},
\end{align}
where the vertical indices remain open, and all other indices are contracted to form the PEPO. It is straightforward to show that the TMs, $\mathbb{H}^{\rm 3TC}$ and $\mathbb{H}^{\rm XC}$, are invariant under the global unitary transformations $U_{ZZ} = \prod_i Z_i \otimes \overline{Z}_i$ and $U_{ZI} = \prod_i Z_i \otimes \overline{I}_i$ using Eqs.\,\eqref{eq:tc-inj-g}, \eqref{eq:3tc_t_tensor_symmetry} and \eqref{eq:xc_t_tensor_symmetry}. It is a simple generalization of Eq.\,\eqref{eq:tc-R-Z-1}. Furthermore, one can show that both TMs, $\mathbb{H}^{\rm 3TC}$ and $\mathbb{H}^{\rm XC}$, also commute with local unitary transformations $u_i^z = Z_i \otimes \overline{Z}_i$ and $u_i^x = X_i \otimes \overline{X}_i$. Similarly, using Eqs.\,\eqref{eq:tc-inj-g}, \eqref{eq:3tc_t_tensor_symmetry}, and \eqref{eq:xc_t_tensor_symmetry}, one can derive the following identities:
\begin{align}
    \includegraphics[width=0.8\linewidth]{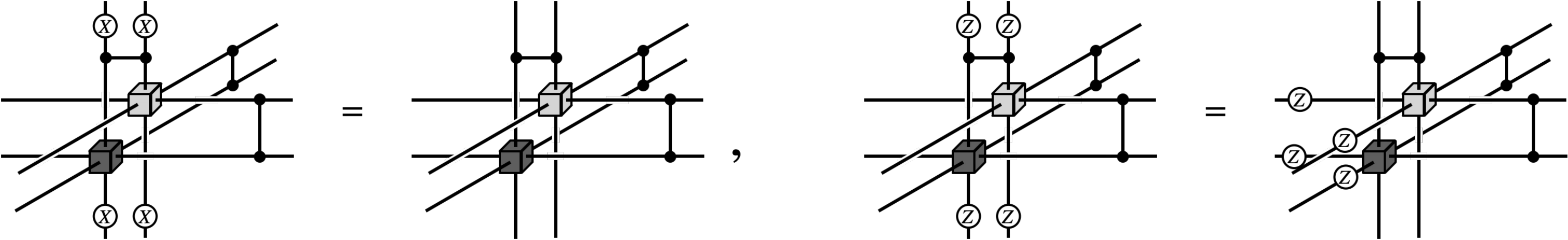}
\end{align}
In the second identity, the Pauli $Z$ appears to remain but cancels out when the tensors are contracted with neighboring ones due to Eq.\,\eqref{eq:tc-inj-g}. In particular, the second identity, or the local symmetry $u_i^z$, remains valid even after the $Z$-filtering operation is applied.

Using Eqs.\,\eqref{eq:tc-inj-g}, \eqref{eq:3tc_t_tensor_symmetry} and \eqref{eq:xc_t_tensor_symmetry}, one can further derive the following equalities for the 3TC state
\begin{align}
    \includegraphics[width=0.6\linewidth]{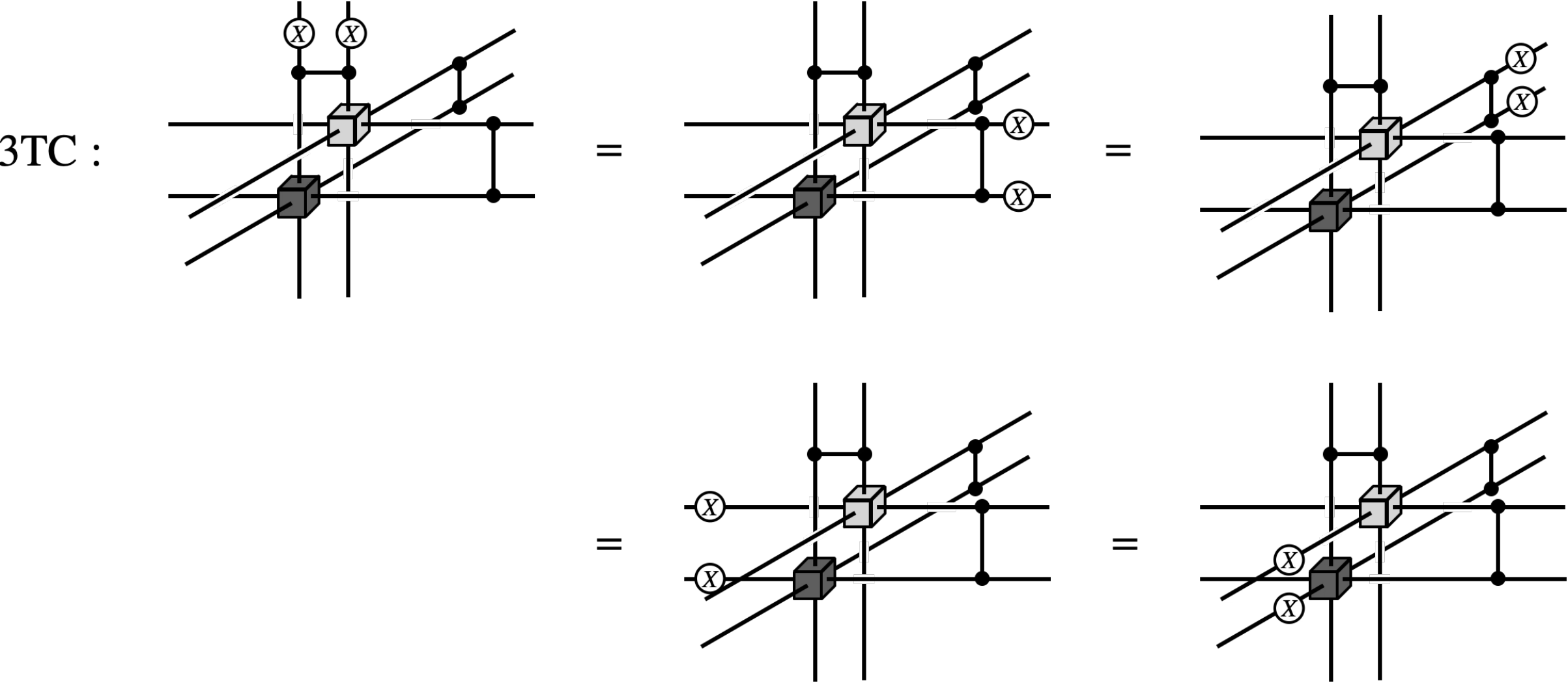},
\end{align}
and for the XC state
\begin{align}
    \includegraphics[width=0.6\linewidth]{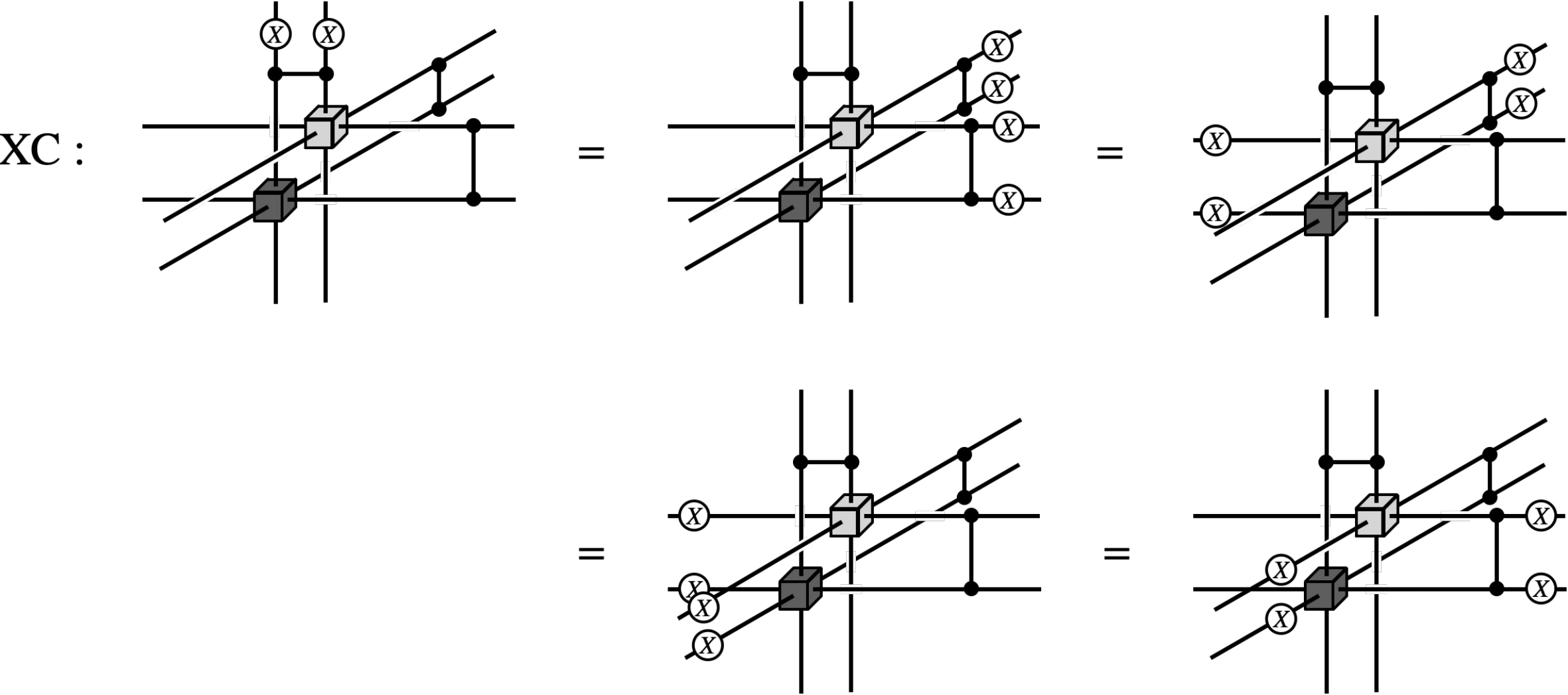}.
\end{align}
Using above equalities, one can verify that $(X_i \otimes \overline{X}_i)(X_j \otimes \overline{X}_j) \mathbb{H}^{\rm 3TC} = \mathbb{H}^{\rm 3TC}$ for all neighboring $i,j$ as graphically illustrated below:
\begin{align}
    \includegraphics[width=0.8\linewidth]{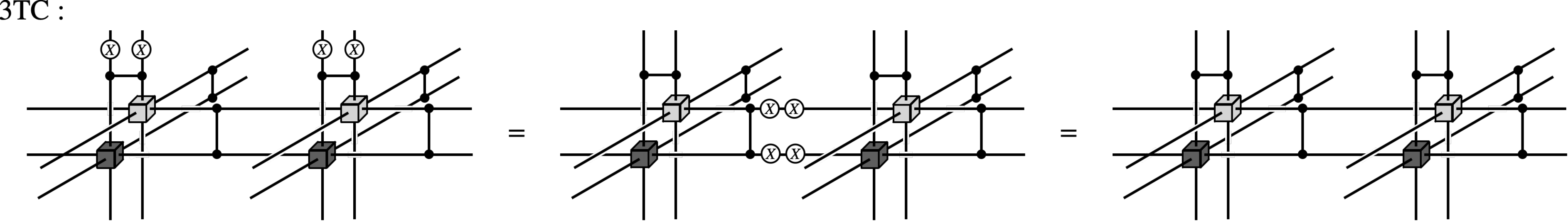},
\end{align}
and $\prod_{i\in\square}(X_i \otimes \overline{X}_i) \mathbb{H}^{\rm XC} = \mathbb{H}^{\rm XC}$ for all elementary plaquettes $\square$ on the square lattice:
\begin{align}
    \includegraphics[width=0.7\linewidth]{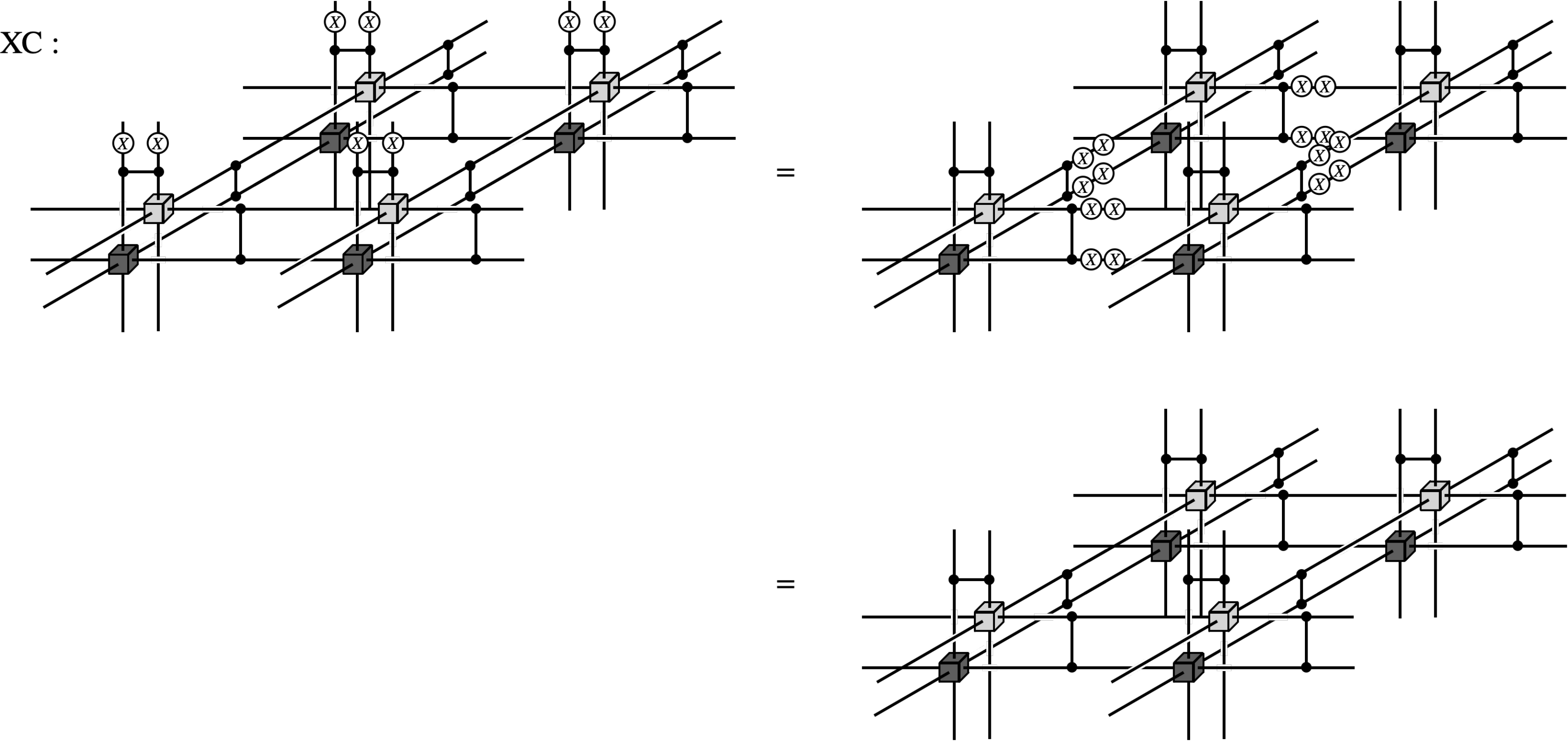}.
\end{align}
These indicate that both TMs are idempotent such that
\begin{align}
    \mathbb{H}^{\rm 3TC} \cong \prod_{\langle ij\rangle} \left[1+ (X_i \otimes \overline{X}_i)(X_j \otimes \overline{X}_j) \right], \quad
    \mathbb{H}^{\rm XC} \cong \prod_{\square} \Big[1+ \prod_{i\in \square} (X_i \otimes \overline{X}_i)\Big].
\end{align}

\end{document}